\documentclass[aps,prd,nofootinbib,reprint,superscriptaddress]{revtex4-2}
\usepackage{hyperref}
\usepackage{graphicx}
\usepackage{overpic}

\usepackage{xcolor}
\usepackage{subcaption}
\usepackage{csquotes}
\usepackage{subcaption}

\usepackage{dcolumn}
\usepackage[normalem]{ulem}
\usepackage{booktabs}
\usepackage{amsmath}

\begin{document}
\preprint{APS/123-QED}
\title{Disentangling the neutron skin of $^{48}$Ca from nuclear radius effects through $^{48}$Ca+$^{48}$Ca and $^{48}$Ti+$^{48}$Ti flow observables}

	\author{Abhisek Saha}
	\affiliation{School of Physics, Peking University, Beijing 100871, China}
	\affiliation{Center for High Energy Physics, Peking University, Beijing 100871, China}

\begin{abstract}
The thickness of the neutron skin ($\Delta r_{np}$) in $^{48}$Ca remains uncertain. CREX and ab initio calculations point to a thin skin, while many energy density functionals prefer larger values. Relativistic heavy-ion collisions can probe the skin through its geometric imprint on the quark‑gluon plasma, but a thicker skin also changes the overall nuclear size, so the two effects are hard to separate. We resolve this by systematically comparing $^{48}$Ca+$^{48}$Ca and $^{48}$Ti+$^{48}$Ti collisions at $\sqrt{s_{NN}}=5.02$ TeV using the AMPT model with string melting. For $^{48}$Ca we take a zero-skin baseline and skins at the CREX lower and upper bounds. As size references we add $^{48}$Ti and a skin-free $^{48}$Ca with the same density parameters as $^{48}$Ti, which separates size from skin effects. The sensitivity to the $^{48}$Ti radius is tested using two choices for its Woods–Saxon radius, one of which reproduces the experimental rms radius. We find that the neutron skin is a purely geometric perturbation; the medium's transport response coefficients are unchanged, first‑moment observables exhibit only modest sensitivity to $\Delta r_{np}$, while second‑moment observables respond much more strongly. The final-state covariance between flow and mean transverse momentum fluctuations, with its normalized correlation, is the most promising measurable probe. The radius-matched $^{48}$Ca follows $^{48}$Ti closely, and the position of $^{48}$Ti relative to the skin configurations depends on its radius. These collisions, analyzed with second-moment and shape-size correlators, offer a promising independent constraint on the $^{48}$Ca neutron skin at the LHC.

\end{abstract}

\keywords{Heaavy-ion collisions, Neutron skin}

\maketitle
\section{Introduction} 			\label{sec:intro}
The nuclear neutron skin thickness $\Delta r_{np}=\langle r^2\rangle_n^{1/2}-\langle r^2\rangle_p^{1/2}$ measures the extent to which the neutron distribution radius extends beyond the proton distribution radius \cite{Reinhard:2010wz}. This difference arises from the isospin asymmetry in neutron‑rich nuclei. The excess neutrons accumulate along the outer surface of the nucleus, forming the neutron skin. It is strongly correlated with the density dependence of the symmetry energy in the nuclear equation of state (EoS) and, hence, with the pressure of neutron-rich matter \cite{Lattimer:2023rpe, Dieperink:2003vs, Chen:2010qx}. While the proton distribution can be measured precisely via electron scattering, the neutron distribution remains difficult to access because neutrons carry no net electric charge.  In $^{48}$Ca, a nucleus with eight excess neutrons, low-energy experiments have provided conflicting constraints. The CREX collaboration reported a thin neutron skin of $0.121\pm0.05$ fm using parity-violating electron scattering  \cite{CREX:2022kgg, Hagen:2015yea, Miyagi:2025rvx}, but with different implementation for the neutron matter equation of state \cite{Yuksel:2022umn}. These results are in line with ab initio calculations from chiral effective field theory ($0.12–0.15$ fm) and coupled‑cluster theory ($0.13-0.16$ fm) \cite{Hagen:2015yea, Simonis:2019spj}.  However, several energy density functionals that describe ground‑state nuclear properties equally well predict significantly larger skins for $^{48}$Ca, up to $\sim 0.22$ fm when constrained by the thick skin of $^{208}$Pb measured by PREX‑II \cite{Yuksel:2022umn, Piekarewicz:2012pp, PREX:2021umo}. No single theoretical framework simultaneously explains the relatively thick skin in $^{208}$Pb measured by PREX-II and the thin skin in $^{48}$Ca seen by CREX. Moreover, a thicker neutron skin is always accompanied by an increase in the overall matter radius so that a change in the nuclear size can mimic a genuine skin effect unless the comparison is carefully controlled.
This situation has created a puzzle in the nuclear physics community. The ambiguity directly affects our understanding of the density dependence of the nuclear symmetry energy, which in turn governs properties of neutron stars and supernovae \cite{Alam:2016cli}.

Heavy-ion collisions at relativistic energies offer a complementary path to constrain the neutron skin. The spatial distribution of nucleons in the colliding nuclei imprints itself on the geometry of the participant region in the initial stage of the collision \cite{Gardim:2011xv, Niemi:2012aj, Gardim:2014tya, Heinz:2013th}. The subsequent collective expansion imprints the geometric shape onto the final-state particle azimuthal anisotropy. The “imaging-by-smashing” technique is used to make a connection between the final-state particle flow and the initial energy deposition and thus predict the nuclear structure of the colliding nuclei \cite{Jia:2025wey, STAR:2024wgy, ALICE:2024nqd, ALICE:2021gxt}. When a neutron-rich nucleus like $^{48}$Ca collides, a diffuse neutron skin enlarges the effective interaction zone and modifies the distribution of participating nucleons. These geometric changes propagate through the evolution of the quark-gluon plasma and ultimately manifest in final-state flow observables. This approach has already proven fruitful. Recent theoretical and experimental efforts have explored these effects in isobar collisions ($^{96}$Ru+$^{96}$Ru vs $^{96}$Zr+$^{96}$Zr) at RHIC \cite{Liu:2022kvz, Xu:2021vpn, Wang:2024mce, Jia:2022qgl}.
Similar sensitivity has been found in Pb+Pb collisions at the LHC, where anisotropic flow measurements show some sensitivity to the neutron skin of $^{208}$Pb, though geometric degeneracies limit the precision \cite{Liu:2023pav}.  Smaller collision systems are particularly promising because the neutron skin constitutes a larger fraction of the nuclear radius, potentially amplifying the relative signal beyond what is achievable in heavy nuclei. The recent oxygen runs at the LHC have further opened the door to such light‑ion studies \cite{ATLAS:2026zgq, ATLAS:2026fos}.

The upcoming $^{48}$Ca+$^{48}$Ca collisions at the LHC offer a special opportunity. But extracting the skin requires a reference system that separates a genuine skin effect from two confounding effects: a change in nuclear size, and a change in participant-number granularity, to which the second-moment observables central to this analysis are directly sensitive. The isobar program at RHIC illustrates the value of a same-mass-number comparison, but $^{96}$Ru and $^{96}$Zr differ in both neutron number and deformation \cite{Jia:2022qgl}. For $^{48}$Ca, a comparison with a same-mass-number reference such as $^{48}$Ti isolates the neutron skin more cleanly.

The more common reference is $^{40}$Ca+$^{40}$Ca. Electron-scattering measurements show that $^{40}$Ca and $^{48}$Ca have nearly identical charge (proton) radii, $\sim 3.38$ fm, despite differing by eight neutrons \cite{GarciaRuiz:2016ohj}. Thus $^{40}$Ca is, in principle, a reasonable proxy for a skin-free $^{48}$Ca. However, the two nuclei differ in mass number, which gives the two collision systems intrinsically different participant-number granularity. The variances and covariances studied here respond to this particle-counting difference independent of any change in nuclear structure, which is precisely the confound that same-mass-number comparisons are designed to avoid \cite{Jia:2022qgl}. A related study \cite{Vitsos:2025jzt} has recently examined the sensitivity of $^{48}$Ca+$^{48}$Ca collective flow to the neutron skin using $^{40}$Ca+$^{40}$Ca as a reference system; the mass-number mismatch should be kept in mind when interpreting second-moment observables in that comparison.

$^{48}$Ti avoids the granularity problem by construction, since it shares the mass number of $^{48}$Ca. It is a stable isotope that has been extensively used in low-energy scattering experiments to determine the charge radius of $^{48}$Ca \cite{Frosch:1968zz}, although it has not previously been collided at the LHC. The experimental rms point-proton radius of $^{48}$Ti is 3.515 fm \cite{Wu:2025sgt}, obtained from the measured charge radius \cite{Frosch:1968zz}, about 0.13 fm larger than that of $^{48}$Ca (3.387 fm). This is comparable to the increase in the matter radius of $^{48}$Ca produced by the range of neutron skins allowed by CREX, so a $^{48}$Ti reference can distinguish a skin effect from a size difference. Moreover, where $^{48}$Ti falls relative to the skin configurations depends on how the Woods-Saxon radius is chosen (Sec.~\ref{secII} and Appendix~\ref{app:ti}).


In this work, we simulate $^{48}$Ca+$^{48}$Ca and $^{48}$Ti+$^{48}$Ti collisions using the AMPT transport model with string melting at $\sqrt{s_{NN}} = 5.02$ TeV. For $^{48}$Ca, we employ three parametrizations of the neutron skin thickness, covering the range of values compatible with existing experimental and theoretical constraints. To separate the size difference from the neutron skin, we additionally construct a skin-free $^{48}$Ca configuration with the same density parameters as $^{48}$Ti. Comparing it with the zero-skin $^{48}$Ca baseline isolates a pure size change; comparing it with $^{48}$Ti tests whether the $Z/N$ composition matters at fixed radius; and comparing it with the skin configurations isolates the effect of the skin at fixed size. We evaluate a comprehensive set of initial-state geometric observables and final-state flow and correlation observables. The goal is to identify which observables carry the clearest imprint of the neutron skin and are measurable at the LHC. The $^{48}$Ti+$^{48}$Ti collisions and the size-matched $^{48}$Ca are used to separate the skin effect from the overall nuclear size. Because this separation depends on the radius assigned to $^{48}$Ti, we also repeat a selection of observables with a larger $^{48}$Ti radius (Appendix~\ref{app:ti}).

The neutron skin acts as a surface‑localised perturbation that primarily modifies the tails of the nucleon distribution. Its influence on event‑averaged quantities is therefore expected to be modest, whereas event‑by‑event fluctuations and correlations (second‑moment observables) may respond more strongly in relative terms. We consequently focus our investigation on variances of the initial size and eccentricity, covariances between shape and size, the event‑plane alignment, and their final‑state analogues. By following how these observables propagate through the partonic and hadronic evolution, we aim to identify signatures of the neutron skin that are experimentally accessible, in view of the upcoming $^{48}$Ca+$^{48}$Ca runs at the LHC.

The paper is organized as follows. Section \ref{secII} describes the AMPT model setup and the three neutron-skin parametrizations used for $^{48}$Ca, together with the $^{48}$Ti and size-matched $^{48}$Ca reference systems. Section \ref{secIII} presents the results in three stages. We first examine how the neutron skin modifies the mean size and shape of the participant zone and their event-by-event fluctuations. We then test whether the resulting flow modifications reflect a change of the medium response or follow from the initial geometry. Finally, we combine shape and size into joint covariance observables, which show the largest sensitivity among the observables studied here. Section \ref{secIV} summarizes our findings and discusses their implications and outlook for future measurements. Appendix \ref{app:angle} presents the event-plane alignment and Appendix \ref{app:ti} repeats a selection of observables with a larger $^{48}$Ti radius.

\section{ Model Setup and Nuclear density Configurations} \label{secII}
We simulate $^{48}$Ca+$^{48}$Ca and $^{48}$Ti+$^{48}$Ti collisions at $\sqrt{s_{NN}} = 5.02$ TeV using the AMPT model with string melting \cite{Lin:2004en}. The model has been widely used to reproduce charged-particle multiplicities and anisotropic flow in heavy-ion collisions from RHIC to LHC energies. The model consists of four stages: initial conditions generation via HIJING \cite{Wang:2000bf}, parton elastic scatterings with Zhang's Parton Cascade (ZPC) \cite{Zhang:1997ej}, hadronization through a quark coalescence mechanism, and hadronic rescatterings via the ART model \cite{Li:1995pra}.

The initial nucleon positions in each incident nucleus are sampled from a Woods--Saxon distribution,
\begin{equation}
    \rho(r) = \rho_0 / (1 + \exp((r - R) / a))
\end{equation}

where $R$ is the half-density radius and $a$ is the diffuseness parameter. A neutron skin is introduced by assigning different half-density radii $R_p$ and $R_n$ to protons and neutrons, with the same diffuseness $a = 0.586$ fm for both species and for all configurations. For this diffuseness the rms radius of each distribution is $\langle r^2\rangle^{1/2} = [\tfrac{3}{5}R^2 + \tfrac{7}{5}\pi^2 a^2]^{1/2}$, so $R$ and the rms radius differ. The neutron skin is defined from the rms radii, $\Delta r_{np} = \langle r^2\rangle_n^{1/2} - \langle r^2\rangle_p^{1/2}$.

Following Ref. \cite{Vitsos:2025jzt}, we use the experimental rms point-proton radius as the input value of $R_p$. For $^{48}$Ca we take $R_p = 3.387$ fm, which equals the rms point-proton radius obtained from electron scattering (3.386-3.388 fm from the 2pF/3pF fits) \cite{Wu:2025sgt}. The point-proton radius is obtained from the charge radius by removing the finite size of the nucleons \cite{Wu:2025sgt}\footnote{AMPT samples point nucleons, so we use point-proton radii and not charge radii. The charge radius is larger because it includes the finite size of the proton. The experimental rms charge radii are $\approx 3.48$ fm for $^{48}$Ca \cite{Angeli:2013epw, Frosch:1968zz} and 3.597 fm for $^{48}$Ti \cite{Wu:2025sgt}, against point-proton radii of 3.387 fm and 3.515 fm.}. Since $R$ is a half-density radius, the sampled rms radius $\langle r^2\rangle_m^{1/2}$ differs from $R$ (Table~\ref{tab:ns}).  A setup that matches the sampled rms to the experimental point-proton rms would require a different $R$ for each nucleus. Because we are interested in a relative comparison with Ref. \cite{Vitsos:2025jzt}, we keep their convention for $^{48}$Ca. configurations. 

For $^{48}$Ti we use two values of the half-density radius $R_p=R_n$, which serve different purposes. In the main text, we use $R_p=R_n=3.515$ fm for $^{48}$Ti. This is the experimental rms point-proton radius of $^{48}$Ti, obtained from the Fourier-Bessel charge density (rms charge radius 3.597 fm) \cite{Wu:2025sgt}. It is used as $R$ in the same way as for $^{48}$Ca, so that the same rule is applied to both nuclei. The sampled rms radius of $^{48}$Ti is then 3.487 fm. In Appendix~\ref{app:ti}, we use $R_p=R_n=3.56$ fm. This value is chosen so that the sampled rms radius, 3.514 fm, reproduces the experimental rms point-proton radius of $^{48}$Ti. The sampled $^{48}$Ca protons have an rms radius of 3.410 fm, which is 0.023 fm above the experimental value. The sampled $^{48}$Ti protons are 0.028 fm below it in the main-text setting and agree with it in the appendix setting. The sampled rms difference between $^{48}$Ti and $^{48}$Ca is therefore 0.077 fm in the main text and 0.104 fm in the appendix, against 0.128 fm in experiment. The two settings differ in the size of the $^{48}$Ti reference, which changes how $^{48}$Ti compares with the neutron-skin configurations. Appendix~\ref{app:ti} tests how the Ca/Ti comparison depends on this choices. The neutron radii $R_n$ of the $^{48}$Ca skin configurations are those of Ref. \cite{Vitsos:2025jzt}. At $a = 0.586$ fm they give rms skins of 0.068 fm and 0.181 fm, the CREX lower and upper bounds. The small neutron skin of $^{48}$Ti is neglected.

\begin{table*}[htbp]
\centering
\caption{Nuclear configurations used in this study. $R_p$ and $R_n$ are Woods--Saxon half-density radii, with $a=0.586$ fm in all cases. The rms radii are those of the sampled distributions, $\langle r^2\rangle^{1/2}=[\tfrac{3}{5}R^2+\tfrac{7}{5}\pi^2a^2]^{1/2}$, with the matter radius $R_{\rm matter} =\langle r^2\rangle_m^{1/2}=[(Z\langle r^2\rangle_p+N\langle r^2\rangle_n)/A]^{1/2}$ and the skin $\Delta r_{np}=\langle r^2\rangle_n^{1/2}-\langle r^2\rangle_p^{1/2}$. For $^{48}$Ca, $R_p=3.387$ fm is the experimental rms point-proton radius. Cases 4 and 5 use $R_p=R_n=3.515$ fm, the experimental rms point-proton radius of $^{48}$Ti. Appendix~\ref{app:ti} uses $R_p=R_n=3.56$ fm for these two cases (sampled rms 3.514 fm).\cite{Loizides:2014vua, Zenihiro:2018rmz, Frosch:1968zz, Chen:2022msv, Angeli:2013epw, Wu:2025sgt}}
\begin{tabular}{c l c c c c c c l}
\toprule
Case & System & $R_p$ (fm) & $R_n$ (fm) & $\Delta r_{np}$ (fm) & $\langle r^2\rangle_m^{1/2}$ (fm) & Description \\
\midrule
1 & $^{48}$Ca+$^{48}$Ca & 3.387 & 3.387 & 0     & 3.410 & No skin; $R_p$ = expt.\ rms point-proton radius of $^{48}$Ca. \\
2 & $^{48}$Ca+$^{48}$Ca & 3.387 & 3.500 & 0.068 & 3.450 & CREX lower bound (thin skin). \\
3 & $^{48}$Ca+$^{48}$Ca & 3.387 & 3.685 & 0.181 & 3.517 & CREX upper bound (thick skin). \\
4 & $^{48}$Ti+$^{48}$Ti  & 3.515 & 3.515 & 0     & 3.487 & Reference: $R_p=R_n$ = expt.\ rms point-proton radius of $^{48}$Ti. \\
5 & $^{48}$Ca+$^{48}$Ca & 3.515 & 3.515 & 0     & 3.487 & No skin; radii matched to $^{48}$Ti (case 4). \\

\bottomrule
\end{tabular}
\label{tab:ns}
\end{table*}

Table~\ref{tab:ns} summarises the five configurations used in this study. The table lists the corresponding proton and neutron Woods–Saxon half-density radii $R_p$ and $R_n$, the neutron skin and the resulting rms nuclear matter radius. In all cases $R_p$ is set equal to the experimental rms point-proton radius, which is a half-density radius here, so the sampled rms differs from it. Cases 1-3 are $^{48}$Ca with three neutron-skin values spanning the CREX range: a zero-skin baseline, a thin skin at the CREX lower bound, and a thick skin at the CREX upper bound. Case 4, the isobaric $^{48}$Ti+$^{48}$Ti system, is the reference. In the main-text setting, the matter rms radius of $^{48}$Ti (3.487 fm) lies between those of Cases 2 and 3 (3.450 and 3.517 fm). With the larger radius of Appendix~\ref{app:ti} it coincides with that of Case 3 to within 0.003 fm. Case 5 is a skin-free $^{48}$Ca configuration with the same density parameters as $^{48}$Ti. It is a radius-matched counterpart of the skin configurations. Comparing it with Case 1 isolates a pure size change, and comparing it with Case 4 tests the dependence on $Z/N$ at fixed radius.

After the initial nucleon positions are set, the string melting version of AMPT converts each nucleon into a quark-antiquark pair or diquark, then proceeds through parton scattering, hadronization, and hadronic rescattering. We keep all model parameters fixed across the five configurations and only vary the nuclear density distributions of the colliding nuclei. The parton scattering cross section is fixed at $\sigma_p = 3$ mb throughout, consistent with values used in prior AMPT studies that reproduce LHC-energy elliptic flow. For each configuration, we generate $\sim 3\times 10^6$ minimum-bias events. The centrality is calculated from the number of charged particles in the pseudorapidity range $|\eta|<0.5$. Statistical uncertainties in all figures are obtained by bootstrap resampling of the event sample within each centrality bin (1000 resamples); the resulting uncertainty bands are propagated through every derived ratio and correlator shown.

We record both the initial-state participant coordinates from which the shape and size of the initial geometry are computed and the final-state charged hadrons within $0.2<p_T<5.0$ GeV/c and $|\eta|<2.5$. All observables presented in the following sections are obtained from the same event sample, allowing direct event-by-event correlations between the initial geometry and the final-state flow. Since the proton distribution is the same in Cases 1-3, any differences among them can be attributed to the neutron skin.


\section{Results} \label{secIII}

\begin{figure*}[t!]
    \centering
    \includegraphics[width=0.48\linewidth]{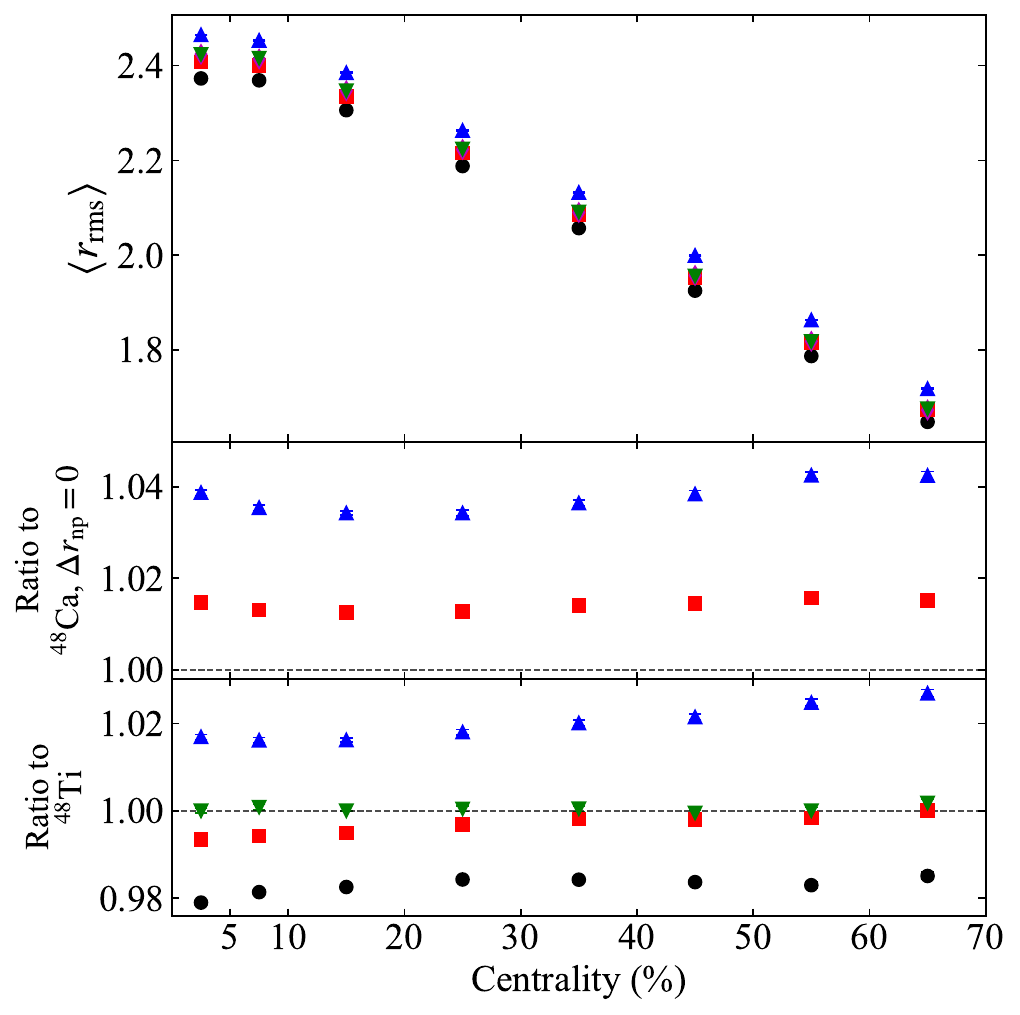}
    \includegraphics[width=0.48\linewidth]{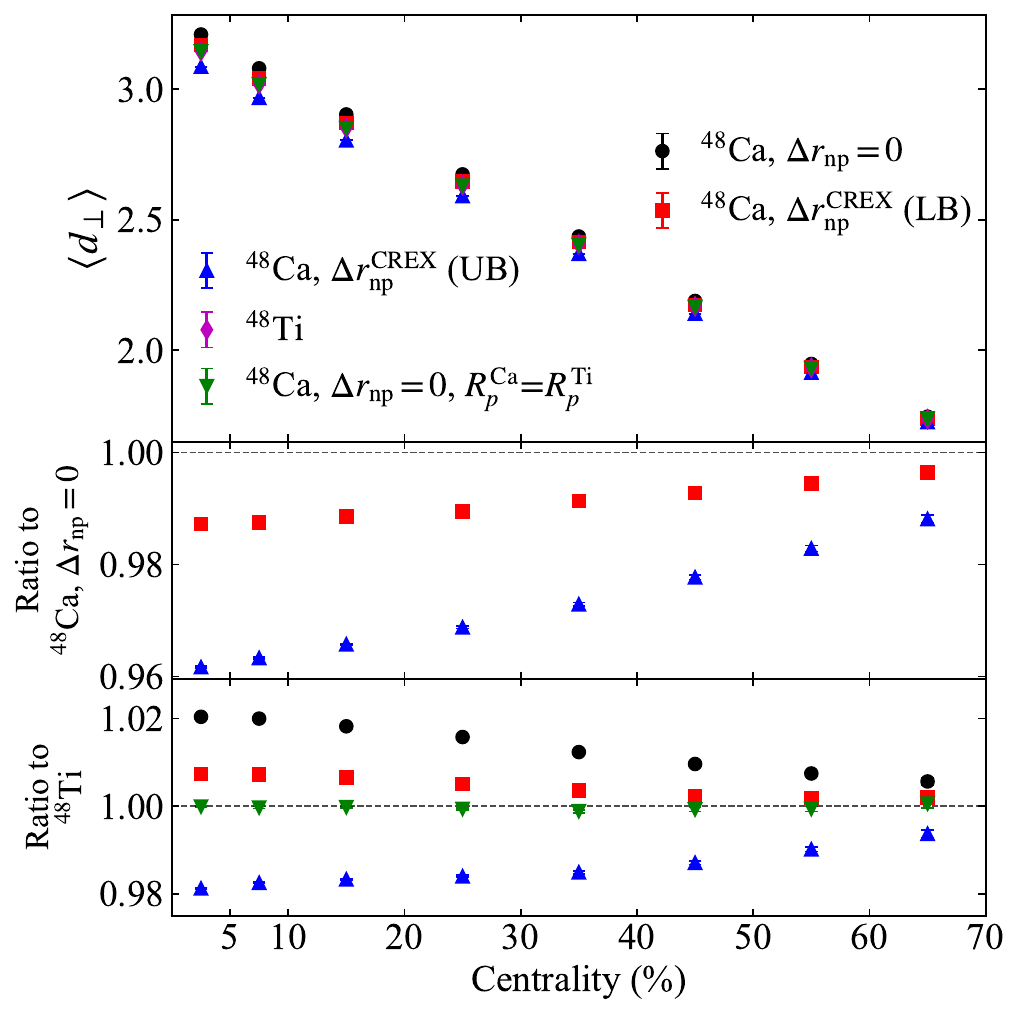}
    \includegraphics[width=0.48\linewidth]{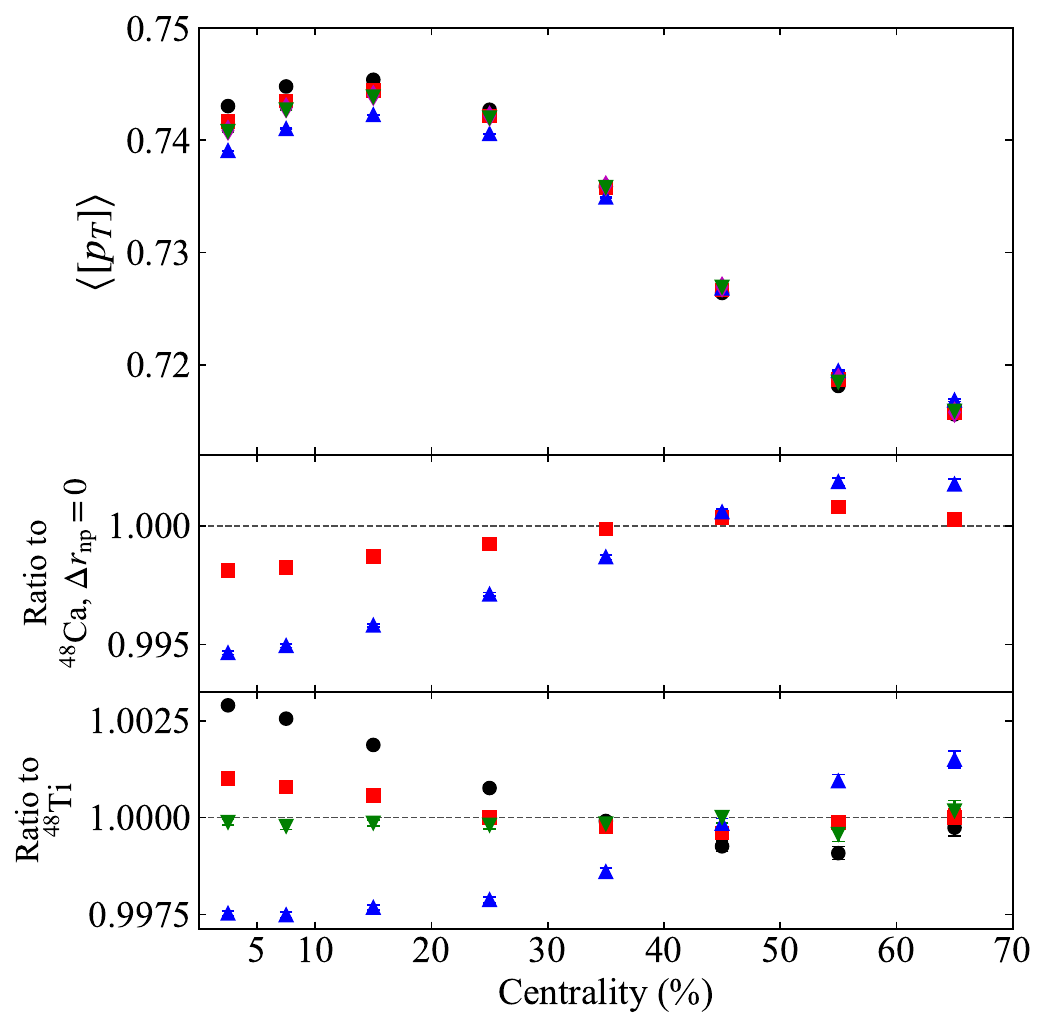}

    \begin{picture}(0,0)
        \put(-200,460){\bfseries (a)}
        \put(50,460){\bfseries (b)}
        \put(-70,200){\bfseries (c)}
        \put(-140, 480){\fontsize{9}{10}\bfseries AMPT\(\boldsymbol{\sqrt{S_{NN}}=5.02}\) TeV}
        \put(10, 233){\fontsize{9}{10}\bfseries \(\boldsymbol{0.2<p_T<3.0}\) GeV/c}
    \end{picture}
    \caption{Centrality dependence of three size-related observables at $\sqrt{s_{\text{NN}}} = 5.02$ TeV from AMPT, for the five nuclear configurations of Table I: the participant RMS radius $r_{\rm rms}$ (top left), mean inverse transverse size $\langle d_\perp \rangle$ (top right), and mean transverse momentum $\langle [p_T] \rangle$ (bottom). Top row: absolute values. Middle row: ratio to the no-skin $^{48}$Ca baseline, isolating the pure neutron-skin effect. Bottom row: ratio to the $^{48}$Ti baseline, testing the isobaric reference strategy. The skin's effect on the mean weakens systematically from the purely geometric $r_{\rm rms}$ to the hydrodynamically-evolved $\langle [p_T] \rangle$, where it even changes sign with centrality}
    \label{fig:fig1}
\end{figure*}

We begin by examining how the neutron skin modifies the geometry of the participant zone before any medium evolution takes place, and how that geometry propagates into final-state flow. To keep the presentation compact, related initial- and final-state observables are grouped into the same figure, so that we can compare them directly. The figures are organized into three-panel layout used throughout this section; the top row shows absolute values for all five configurations, the middle row isolates the pure neutron-skin effect by normalizing to the no-skin $^{48}$Ca baseline ($\Delta r_{np}=0$), and the bottom row provides the isobaric test by normalizing to the $^{48}$Ti reference. This convention is used for every figure in this section.
Throughout this section, the following color and symbol coding is used uniformly; the no‑skin $^{48}$Ca baseline is shown as black circles; the CREX lower bound ($\Delta r_{np}=0.068$) as red squares; the CREX upper bound ($\Delta r_{np}=0.181$) as blue triangles; the matched-radius $^{48}$Ca as green downward‑pointing triangles; and the $^{48}$Ti reference as purple diamonds. The centrality range shown varies between figures: mean and variance observables (Figs. \ref{fig:fig1}-\ref{fig5_kappa}) are shown out to 60-70\% centrality, while the shape-size covariance and its normalized form (Figs. \ref{fig6_cov}-\ref{fig7_rho2}) are restricted to 40\%, where the three-subevent method retains sufficient per-event multiplicity for a stable covariance estimate.

\subsection{Size and momentum: means and their fluctuations}

Figure \ref{fig:fig1} shows three size-related quantities as a function of centrality, the mean participant RMS radius $\langle r^2\rangle^{1/2}_{\text{part}}$ (panel a), the mean inverse transverse size $\langle d_{\perp} \rangle$ (panel b), and the mean transverse momentum $\langle p_{T} \rangle$ (panel c). Together they trace how a purely geometric perturbation at the nucleon level propagates to the initial density gradient and then to the final-state radial flow. The three quantities respond to the skin with markedly different sensitivity.

$\langle r_{rms} \rangle$ is a direct measure of the absolute size of the interaction region and responds most strongly. Relative to the no-skin baseline, a thicker skin inflates $r_{rms}$ by $\sim 3.5–4\%$ for the CREX upper bound (UB) and by $\sim 1.5-1.8\%$ for the lower bound (LB) (Fig. \ref{fig:fig1}a, middle row). This fractional increase is nearly independent of centrality over the 5–65\% range. This is what one expects from a static, surface-localized perturbation. The extra neutrons sit at the nuclear surface, and their contribution scales with the number of surface nucleons that participate, not with the impact parameter.

While $r_{\rm rms}$ measures the absolute spatial extent of the participant zone, $d_{\perp}=\sqrt{N_{\rm part}/S_{\perp}}$ provides a density-weighted inverse length scale that better tracks the initial pressure-gradient scale driving radial flow, making it a more direct geometric proxy for $[p_T]$ \cite{Bhalerao:2005mm, Jia:2021qyu}. Here $S_{\perp}=\pi R_{\perp}^2=\pi\sqrt{\langle x^2\rangle \langle y^2\rangle}$ is the transverse overlap area and $N_{\rm part}$ is the number of participants. Within a fixed centrality bin, a thicker skin pushes neutrons outward into a diffuse shell. This inflates the transverse overlap area $S_\perp$  without appreciably changing the number of participants and so suppresses $\langle d_\perp \rangle$ (Fig. \ref{fig:fig1}b). The suppression is $\sim$4\% for UB and $\sim$1.5\% for LB in central collisions. It decreases monotonically with centrality, to $\sim$2\% for UB at 65\%, because in peripheral events the participant zone shrinks and the skin’s fractional contribution to the overlap area becomes small.

$\langle [p_T]\rangle$ encodes the time-integrated radial flow built up hydrodynamically from the initial gradient, and is the least sensitive of the three.
In central collisions the skin suppresses $\langle[p_T]\rangle$ by only $\sim$0.55\% for UB and $\sim$0.2\% for LB and the effect reverses toward peripheral centralities, turning into a very small enhancement (Fig. \ref{fig:fig1}c). This sign change is absent in the purely geometric $\langle d_\perp \rangle$ and only appears once the medium evolution is included. A plausible interpretation is a competition between two effects. A larger skin dilutes the initial pressure gradient, which suppresses $[p_T]$. It also enlarges the system and prolongs the fireball lifetime, which allows more time for radial flow to build up and enhances $[p_T]$. In central collisions gradient dilution dominates. In peripheral collisions, where the gradient difference between configurations is small to begin with, the longer lifetime becomes the controlling factor. A similar trend has been reported in earlier studies \cite{Vitsos:2025jzt}. We note that we have not isolated these two contributions separately in AMPT.

The bottom row of Figure \ref{fig:fig1} shows the isobaric test central to our reference strategy. For all three observables, the $R_p$-matched $^{48}$Ca configuration (green downward triangles) reproduces the $^{48}$Ti baseline to within $\sim$0.1\% for $\langle r_{\rm rms}\rangle$ and $\langle d_\perp\rangle$ and $\sim$0.05\% for $\langle[p_T]\rangle$, across the whole centrality range. 
This confirms that matching the effective proton radius of $^{48}$Ca to that of $^{48}$Ti removes the gross nuclear-size difference, and the mean observables of $^{48}$Ti depend on its density profile essentially through $R_p$. The physically smaller no-skin $^{48}$Ca baseline sits below the $^{48}$Ti reference in size by $\sim$2\% in $\langle r_{\rm rms}\rangle$, and by 2.0\% above it in $\langle d_\perp\rangle$. In $\langle[p_T]\rangle$ it is $\sim$0.3\% above $^{48}$Ti in central collisions.

Because the sampled rms radius of $^{48}$Ti (3.487 fm) lies midway between the matter radii of the LB (3.450 fm) and UB (3.517 fm) scenarios, the CREX range brackets the $^{48}$Ti reference. The LB scenario lies closest to $^{48}$Ti in all three observables. The UB scenario overshoots it, by $2$–$2.5\%$ in $\langle r_{\rm rms}\rangle$, $-2\%$ in $\langle d_\perp\rangle$ and $-0.25\%$ for $\langle[p_T]\rangle$ in central collisions. In $\langle r_{\rm rms}\rangle$ the $^{48}$Ti reference is therefore equivalent to a $^{48}$Ca skin of $\Delta r_{np}\approx0.07$–0.10 fm, depending on centrality, i.e. in the lower half of the CREX range. The reason is that a skin adds a diffuse surface layer, whereas the $R_p$-matched control rescales the whole density profile, so $R_{\rm matter}$ alone does not fix the participant-zone size. A deviation of a measured Ca/Ti ratio from unity in a size-like observable thus indicates whether the true skin is thicker or thinner than this intermediate value. For this reason we use the $^{48}$Ti reference, together with the $R_p$-matched control, to separate skin from size effects. We do not repeat this argument for every subsequent figure; where the same $R_p$-matched collapse appears below, it should be read as a repetition of this test.

\subsection{Fluctuations of size and momentum}

\begin{figure*}[t!]
    \centering
    \includegraphics[width=0.45\linewidth]{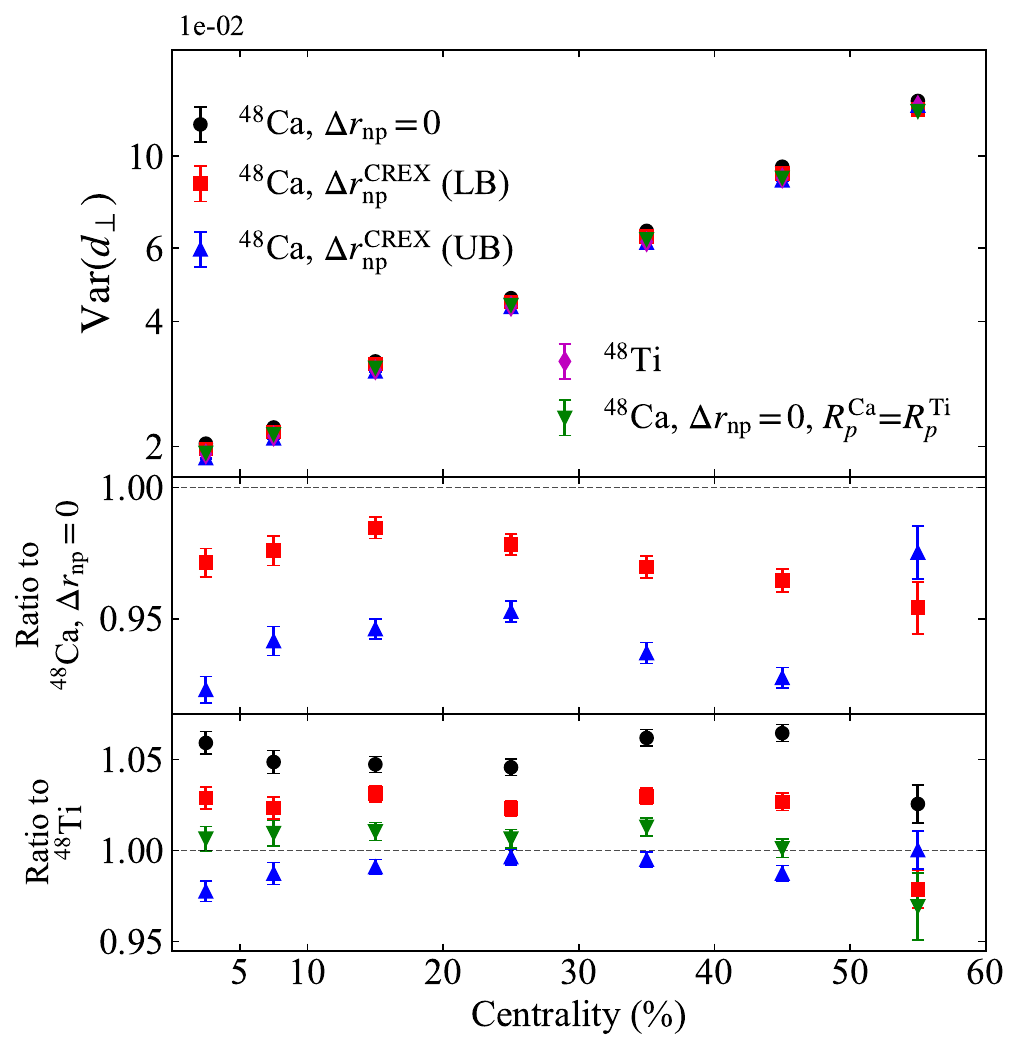}
    \includegraphics[width=0.45\linewidth]{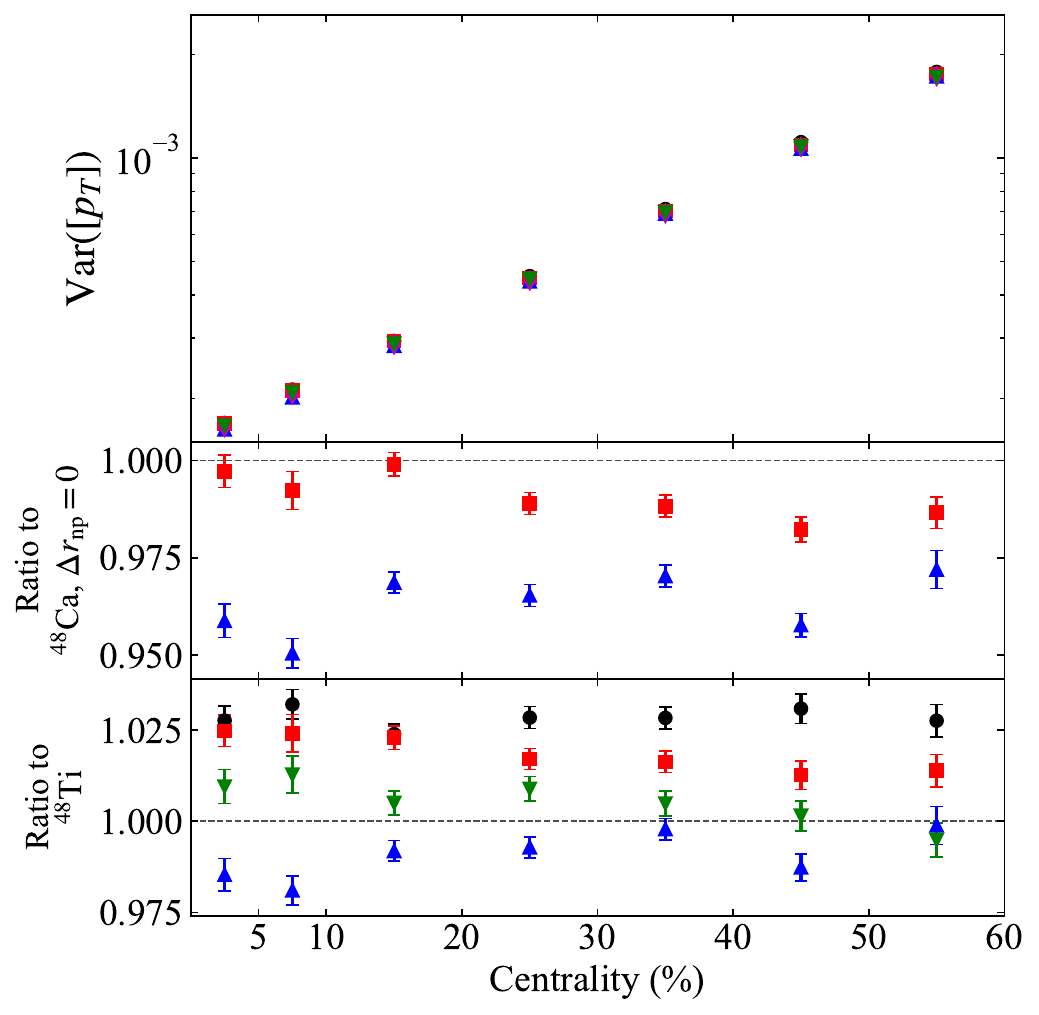}
    \begin{picture}(0,0)
        \put(-425,215){\bfseries (a)}
        \put(-185,213){\bfseries (b)}
        \put(-160, 210){\fontsize{9}{10}\bfseries AMPT\(\boldsymbol{\sqrt{S_{NN}}=5.02}\) TeV}
        \put(-110, 135){\fontsize{9}{10}\bfseries \(\boldsymbol{0.2<p_T<3.0}\)}
    \end{picture}
    \caption{Centrality dependence of the event-by-event variances of the inverse transverse size, $\mathrm{Var}(d_{\perp})$ (left), and of the mean transverse momentum, $\mathrm{Var}([p_T])$ (right), for the five nuclear configurations. Panel layout as in Fig. \ref{fig:fig1}. The neutron skin suppresses both variances, with a magnitude roughly twice as large in the initial state ($\mathrm{Var}(d_{\perp})$) as after medium evolution ($\mathrm{Var}([p_T])$), consistent with the skin acting on the width of the underlying distributions rather than their means.}
    \label{fig2:var_dp_mpt}
\end{figure*}

Figure \ref{fig2:var_dp_mpt} shows the event-by-event variances $\mathrm{Var}(d_{\perp})$ (panel a) and $\mathrm{Var}([p_T])$ (panel b), the fluctuation counterparts of the two hydrodynamically-linked quantities in Fig. \ref{fig:fig1}. $\mathrm{Var}(d_{\perp})$ characterises the event-by-event fluctuations of the initial‑state density gradient and it reflects how broadly the system size fluctuates from event to event, while $\mathrm{Var}([p_T])$ is its final-state analogue. As the number of participants drops and event-to-event fluctuations become more significant, both variances increase from central to peripheral collisions. The no-skin baseline has the largest variances and the CREX upper bound (UB) the smallest, since a more compact system has larger relative fluctuations of its density gradient. 

Relative to the no-skin baseline (middle row), the skin suppresses $\mathrm{Var}(d_{\perp})$ by $\sim3-5\%$ for CREX lower bound and $\sim5–8\%$ for CREX upper bound, with the suppression strongest in the central collisions and persist across all centralities. $\mathrm{Var}([p_T])$ is suppressed by $\lesssim$2\% for LB ( consistent with unity in some bins) and by $\sim$3–5\% for UB.
The origin of the $\mathrm{Var}(d_\perp)$ suppression is largely a rescaling.  At fixed $N_{\rm part}$, the skin inflates the mean overlap area $S_\perp$ and lowers $\langle d_\perp\rangle$ by $\sim$4\% (UB) in central collisions (Fig. \ref{fig:fig1}b). This accounts for the larger suppression of $\mathrm{Var}(d_\perp)$. The skin therefore shifts the scale of the $d_\perp$ distribution without changing its relative width. 
This is the first indication that fluctuation observables carry a stronger skin signature than their corresponding means, a pattern that recurs throughout this section.

$\mathrm{Var}([p_T])$ shows the same qualitative suppression but at slightly smaller in magnitude of $\mathrm{Var}(d_{\perp})$. This damping is consistent with hydrodynamic expansion acting as a low-pass filter on the initial density-gradient fluctuations, and thus only the fraction of the initial variance that is linearly transmitted through the time-integrated radial flow survives into the final state, making $\mathrm{Var}([p_T])$ a cleaner but less amplified probe than its initial-state counterpart. 

The bottom panel shows $\mathrm{Var}(d_{\perp})$ and $\mathrm{Var}([p_T])$ normalized to the $^{48}$Ti+$^{48}$Ti baseline. The $R_p$-matched $^{48}$Ca control lies very close to unity. The physically smaller no-skin baseline lies above $^{48}$Ti by $\sim$5–6\% in $\mathrm{Var}(d_\perp)$ and $\sim$2.5–3\% in $\mathrm{Var}([p_T])$. This also show that a more compact system has larger fluctuations. The $^{48}$Ti variances are bracketed by the two CREX scenarios. The UB lies below $^{48}$Ti by $\sim$2.5\% at the most central bin, and the gap closes to $\lesssim$0.5\% at 25–35\%. This confirms $R_{\rm matter}$ as the controlling parameter.

\subsection{Shape and its flow response}

\begin{figure*}[t!]
    \centering
    \includegraphics[width=0.48\linewidth]{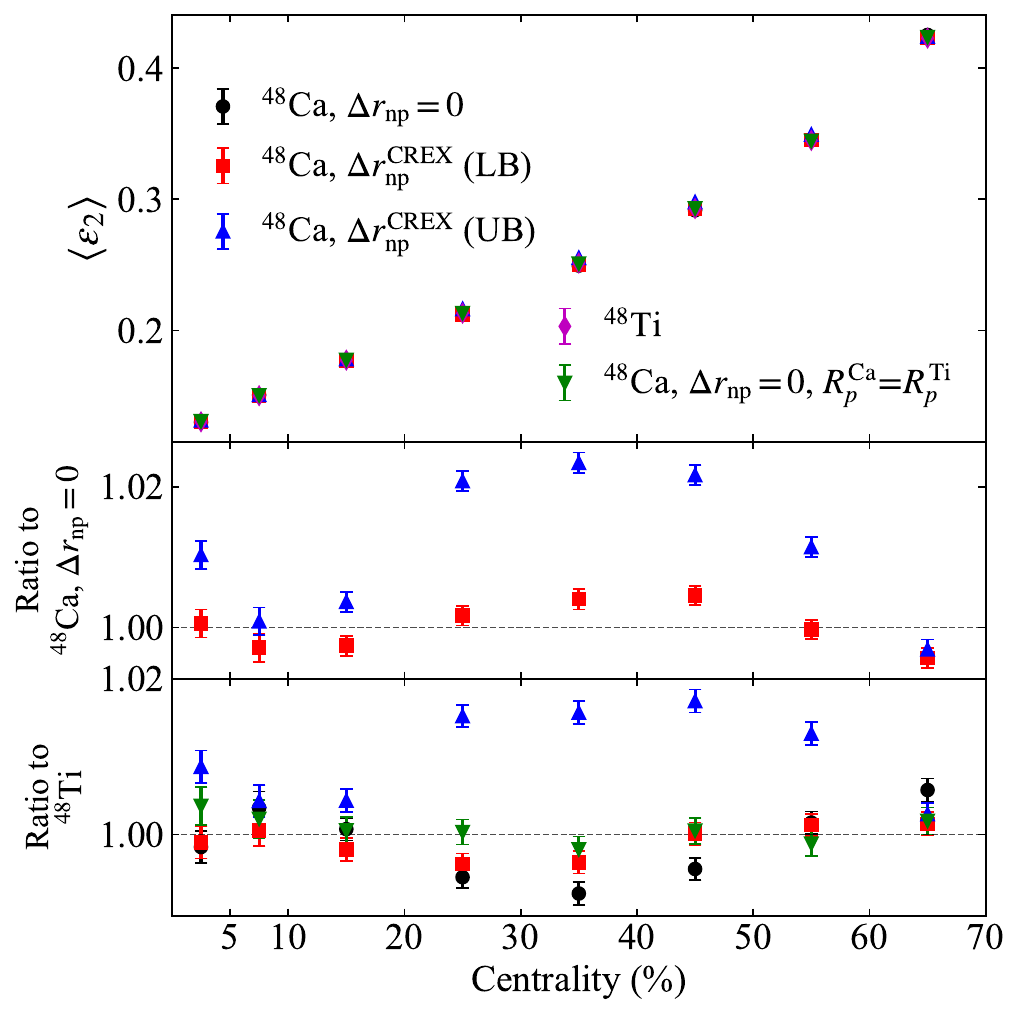}
    \includegraphics[width=0.48\linewidth]{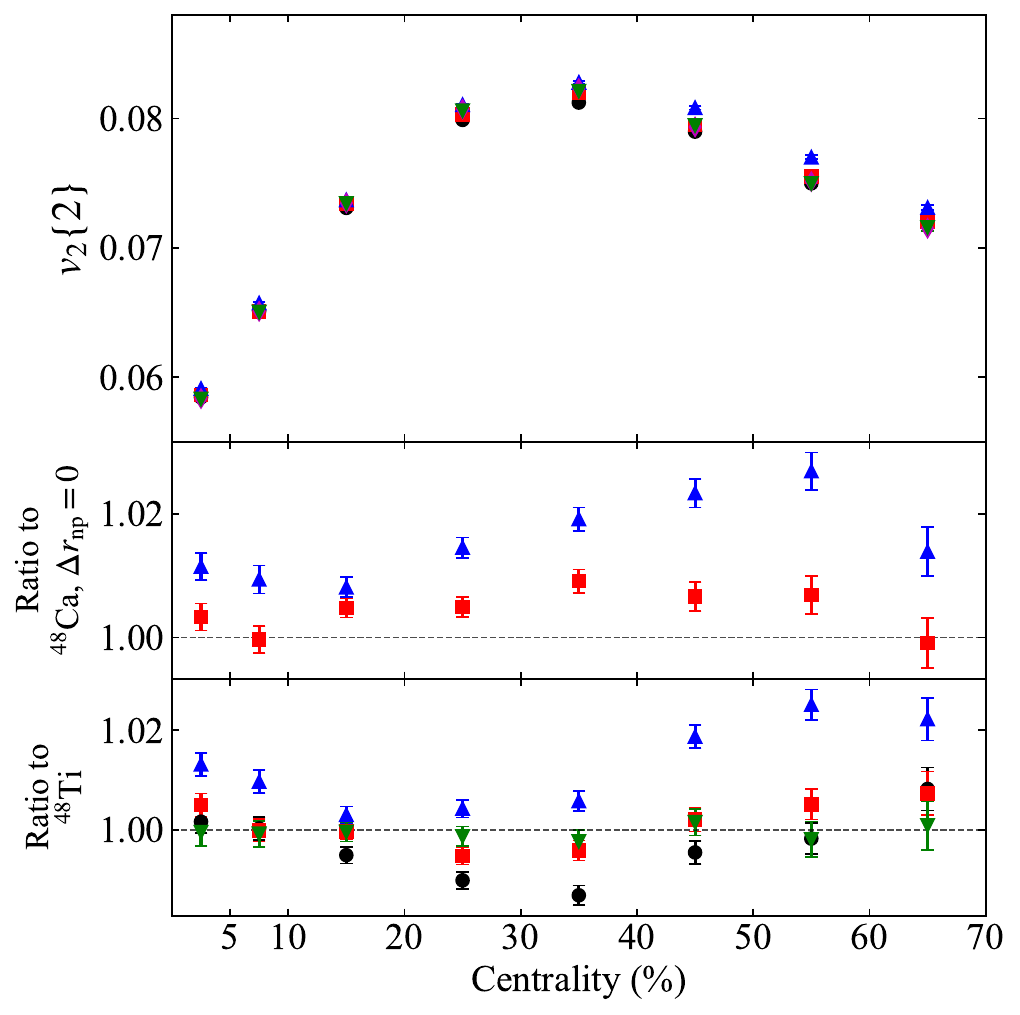}
    \begin{picture}(0,0)
        \put(-435,230){\bfseries (a)}
        \put(-195,230){\bfseries (b)}
        \put(-160, 230){\fontsize{9}{10}\bfseries AMPT\(\boldsymbol{\sqrt{S_{NN}}=5.02}\) TeV}
        \put(-110, 150){\fontsize{9}{10}\bfseries \(\boldsymbol{0.2<p_T<3.0}\)}
    \end{picture}
    \caption{Centrality dependence of the mean eccentricity $\langle\varepsilon_2\rangle$ (left) and the mean elliptic flow $v_2\{2\}$ (right), for the five nuclear configurations. Panel layout as in Fig. \ref{fig:fig1}. The skin's enhancement of $v_2\{2\}$ closely tracks its enhancement of $\langle\varepsilon_2\rangle$ in both shape and magnitude, consistent with a purely geometric origin propagated through an unmodified medium response.}
    \label{fig3_eps2_v2}
\end{figure*}

We now turn from the size of the participant zone to its shape, and pair each shape observable directly with its final-state flow analog. Figure \ref{fig3_eps2_v2} shows the mean eccentricity $\langle \varepsilon_2\rangle$ (panel a) and the mean elliptic flow $v_2\{2\}$ (panel b). The eccentricity $\varepsilon_2$ quantifies the shape of the overlap area \cite{Jia:2021qyu},
\begin{equation}
    \varepsilon_2 e^{i2\Phi_2}=-\frac{\langle r_{\perp}e^{i2\phi}\rangle}{\langle r_\perp \rangle}
\end{equation}
Here the average runs over participant nucleons in the transverse plane $(x,y) =(r_\perp, \phi)$ and the coordinates are rotated into the centre‑of‑mass frame such that x (y) corresponds to the minor (major) axis, quantifying the ellipticity of the initial‑state geometry.

$\langle \varepsilon_2\rangle$ increases monotonically from $\sim 0.1$ in central to $\sim 0.4$ in peripheral collisions as also observed in other spherical nuclei such as $^{197}$Au\cite{Bhalerao:2005mm, Qiu:2011iv}. In contrast to the size observables of Fig. \ref{fig:fig1}, all five configurations lie almost on top of each other in the absolute panel. Relative to the no-skin baseline ( Fig. \ref{fig3_eps2_v2}a, middle row), the lower bound (LB) changes $\langle\varepsilon_2\rangle$ by at most $\sim$0.5\%, while the upper bound (UB) enhances it by up to $\sim$2.4\%. The UB enhancement is non-monotonic in centrality. It is $\sim$1\% in the most central bin rises to a maximum of $\sim$2.4\% at 35\% and falls back to $\sim$1.1\% at 55\%. This near-degeneracy reflects the fact that, at fixed $N_{\rm part}$, the mean shape of the overlap is set predominantly by the impact parameter rather than by the radial density profile. A plausible origin of the mid-central enhancement is centrality-bin migration. A thicker skin increases $N_{\rm part}$ at fixed impact parameter $b$, so selecting the same $N_{\rm part}$-defined centrality bin requires a slightly larger $b$, which is intrinsically more eccentric. This effect peaks at 30–40\% centrality, where the $\varepsilon_2$-$b$ relation is steepest, and fades in both the central and very peripheral limits. Anyhow, it does not account for the $\sim$1\% UB enhancement in the most central bin, and we have not investigated that bin separately.


The final-state anisotropic flow is obtained with the two-subevent cumulant method, which suppresses non-flow correlations through a $\Delta\eta$ gap \cite{Zhou:2015iba}. The subevents $A$ and $B$, separated in pseudorapidity, have multiplicities $M_A$,$M_B$ and flow vectors $Q_2^{A,B}=\sum_{j\in A,B} e^{i2\phi_j}$ for each event. The per-event correlation and its multiplicity weight are,
\begin{equation}
    c_2^{(k)}=\frac{{\rm Re}\big[Q_2^{A}(Q_2^{B})^{*}\big]_k}{M^k_AM^k_B},\qquad w_k=M^k_AM^k_B, 
\end{equation}
where $k$ runs over events. The multiplicity-weighted average is essential for correctly accounting for the event-by-event statistical fluctuation,

\begin{equation}
    \langle\langle 2\rangle\rangle=\frac{\sum_k w_k c_2^{(k)}}{\sum_k w_k},\qquad v_2\{2\}=\sqrt{\langle\langle 2\rangle\rangle}. 
    \label{eq:v2cum}
\end{equation}
$v_2\{2\}$ mirrors the $\langle\varepsilon_2\rangle$ hierarchy but is non-monotonic in centrality. It rises to a broad maximum of $\sim 0.082$ around 30–40\% before falling to $\sim 0.072$ at 65\%. This reflects the competing effect of a shrinking, shorter-lived fireball at large impact parameter, which limits how efficiently eccentricity is converted into momentum-space flow \cite{Song:2010mg, Molnar:2001ux, Romatschke:2007mq}. The skin-induced enhancement ($\sim 0.5–1\%$ for the lower bound, $\sim1–2\%$ for the upper bound relative to $\Delta r_{np}=0$) closely tracks the $\langle \varepsilon_2 \rangle$ enhancement in shape and magnitude, but its peak is shifted to somewhat more peripheral centralities (40–60\%), reflecting the additional centrality dependence of the linear response coefficient $\kappa= v_2\{2\}/\varepsilon_2$ discussed in Sec. III.B \cite{Song:2010mg}. This close correspondence is itself evidence that the skin's effect on $v_2\{2\}$ is inherited directly from the initial geometry rather than from any change in the medium response, a point we verify explicitly in the next section.

The bottom rows of Fig. \ref{fig3_eps2_v2} show that the $R_p$-matched $^{48}$Ca control reproduces $^{48}$Ti to within $\sim$0.3\% in both $v_2\{2\}$ and $\langle\varepsilon_2\rangle$ across the whole centrality range. The $^{48}$Ti reference separate the zero skin and the upper bound level, the LB lies $\sim$0.5–1.0\% below it and the no-skin baseline $\sim$1–2\% below it. The CREX upper bound lies close to $^{48}$Ti in central collisions and overshoots in the peripheral collisions at 45–65\% by $\sim$2.5\%. 

\subsection{Shape fluctuations and flow variance}
\begin{figure*}[t!]
    \centering
    \includegraphics[width=0.48\linewidth]{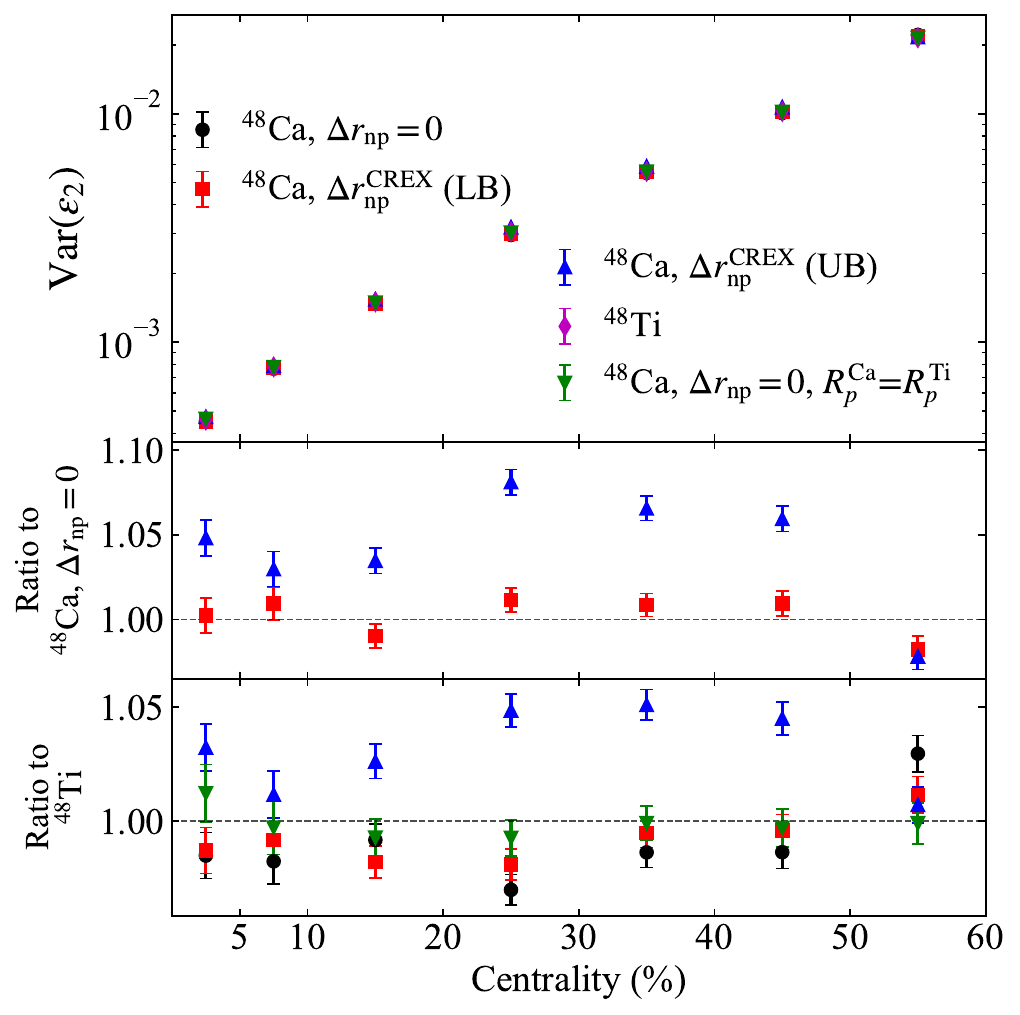}
    \includegraphics[width=0.48\linewidth]{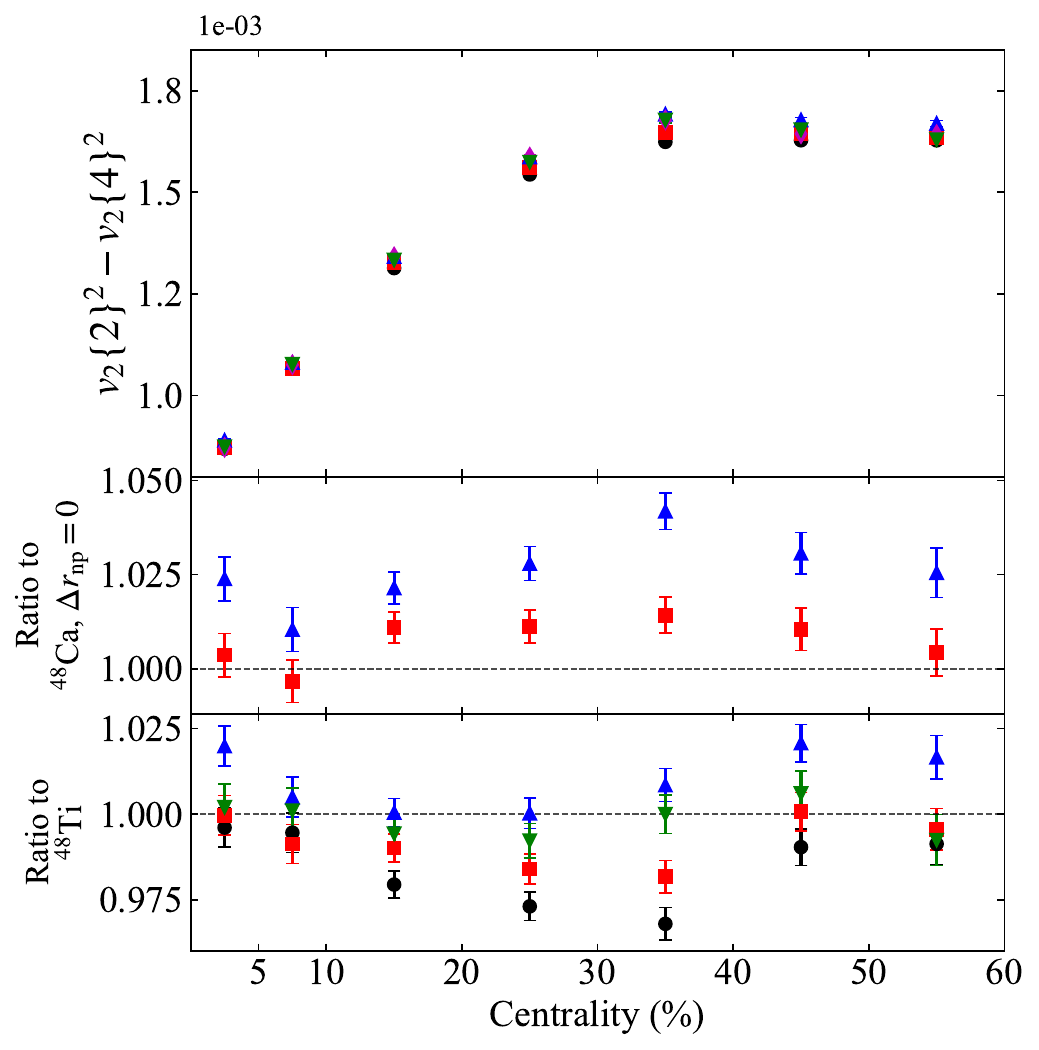}
    \begin{picture}(0,0)
        \put(-445,230){\bfseries (a)}
        \put(-200,225){\bfseries (b)}
        \put(-160, 225){\fontsize{9}{10}\bfseries AMPT\(\boldsymbol{\sqrt{S_{NN}}=5.02}\) TeV}
        \put(-110, 150){\fontsize{9}{10}\bfseries \(\boldsymbol{0.2<p_T<3.0}\)}
    \end{picture}
    \caption{Centrality dependence of the eccentricity variance $\mathrm{Var}(\varepsilon_2)$ (left) and the flow variance $\mathrm{Var}(v_2)$ (right), for the five nuclear configurations. Panel layout as in Fig. \ref{fig:fig1}. Both variances carry a larger skin signature than their respective means (Fig. \ref{fig3_eps2_v2}); the flow variance separates between configurations earlier in centrality than the eccentricity variance, pointing to a residual amplification introduced by the medium response.}
    \label{fig4_var_eps2_v2}
\end{figure*}

Although the variance of $v_2$ ($\mathrm{Var}(v_2)$) cannot be measured directly, it can be obtained experimentally using multi-particle cumulants of $v_2$ as  \cite{STAR:2011ert, Yi:2012gx, ALICE:2016kpq},
\begin{equation}
    v_2\{2\}^2 = \langle v_2^2 \rangle + \sigma^2_{v_2}+\delta_2, \qquad v_2\{4\}^2 \simeq  \langle v_2^2 \rangle - \sigma^2_{v_2}
    \label{eq:v2diff}
\end{equation}
Thus, the difference between the two directly isolates $\sigma_{v_2}^2=\mathrm{Var}(v_2)$ up to the non-flow term $\delta_2$. Unlike $\mathrm{Var}([p_T])$ and $\mathrm{Var}(v_2)$, which are measured, $\mathrm{Var}(d_\perp)$ and $\mathrm{Var}(\varepsilon_2)$ are initial-state quantities that are not directly observable. They are constrained through the assumed linear medium response \cite{Gardim:2011xv, Bozek:2012fw} or through global Bayesian calibration of initial-state models to final-state flow and $[p_T]$ data \cite{Bernhard:2016tnd, Soeder:2023vdn}.


Figure \ref{fig4_var_eps2_v2} shows $\mathrm{Var}(\varepsilon_2)$ (left) and $\mathrm{Var}(v_2)$ (right). Both increase with centrality, $\mathrm{Var}(\varepsilon_2)$ from $\sim 4\times 10^{-4}$ to $\sim 2\times 10^{-2}$, $\mathrm{Var}(v_2)$ from $\sim 8\times 10^{-4}$ to $\sim 1.6\times 10^{-3}$, reflecting growing eccentricity and flow fluctuations toward peripheral collisions. The fluctuations of $v_2$ measured by the multi-particle cumulant approach employing Eq. \ref{eq:v2diff} in ref. \cite{STAR:2011ert} exhibit a similar pattern. As for $\langle\varepsilon_2\rangle$ and $v_2\{2\}$, the five configurations are nearly indistinguishable in the top panels, because at fixed $N_{\rm part}$ the fluctuations are driven mainly by the collision geometry rather than by the fine details of the density profile.

The skin signal is clearer in the ratios. Relative to the no-skin baseline, $\mathrm{Var}(\varepsilon_2)$ is enhanced by 0-1.5\% (lower bound) to 2–7\% (upper bound), peaking in mid-central collisions before collapsing toward unity at 60\% centrality. This is few times the UB enhancement of $\langle\varepsilon_2\rangle$ (Fig. 3a) but follows the same order and shape. A plausible reason is that centrality-bin migration not only shifts the mean $\varepsilon_2$ upward; it also broadens the $\varepsilon_2$ distribution by drawing more eccentric events from larger impact parameters, and $\mathrm{Var}(\varepsilon_2)$ is sensitive to both effects at once. $\mathrm{Var}(v_2)$ shows the same pattern with a smaller amplitude: $\lesssim1.4\%$ for the LB and 1–4\% for the UB, peaking at 35\%. The isobaric test (bottom row) differs between the two observables but it is similar to what was seen in $\varepsilon_2$ and $v_2$. 


\subsection{ Medium response tests: geometry‑to‑flow coupling} \label{sec:medium}

The mean and variance analysis showed that the neutron skin modifies both the size and the shape of the overlap region. The skin inflates the transverse area $S_\perp$,  thereby suppressing $\langle d_\perp \rangle$ and $\mathrm{Var}(d_\perp)$, while simultaneously enhancing  $\langle \varepsilon_2\rangle$ and $\mathrm{Var}(\varepsilon_2)$ through centrality bin migration. These geometric changes propagate into the final-state flow observables $v_2\{2\}$ and $\mathrm{Var}(v_2)$ with the same ordering and $R_{\rm matter}$ scaling. However, a quantitative ambiguity remains. The Ca/Ti deficit from the initial density distribution to final particle flow could originate either from a restructuring of the initial geometry or from a genuine modification of the medium’s dynamical response when the nuclear density profile changes \cite{Gardim:2014tya, Gardim:2011xv}. We separate the two with two questions. First, does the skin change the mean transport response, as measured by $\kappa=\langle v_2\rangle/\langle\varepsilon_2\rangle$? Second, does it change the event-by-event coupling between geometry and flow? For the second question we use three independent statistics of the same $v_2-\varepsilon_2$ joint distribution: the regression slope $\kappa_{\rm slope}$ \big($=\frac{\mathrm{Cov}(v_2, \varepsilon_2)}{\mathrm{Var}(\varepsilon_2)}$\big), the Pearson correlation $\rho(v_2, \varepsilon_2)$, and the event-plane alignment $\langle \cos[2(\Psi_2^{\rm init}- \Psi_2^{\rm final})]\rangle$ \cite{Gardim:2011xv, Niemi:2012aj, Qiu:2012uy}. Here $\Psi_2^{\rm init}$ is the participant‑plane angle obtained from the initial $r^2$-weighted eccentricity vector of the participant nucleons, while $\Psi_2^{\rm final}$ is the azimuthal angle of the final‑state flow vector.
We present $\kappa$, $\kappa_{\rm slope}$, and $\rho(v_2,\varepsilon_2)$ together in Fig. \ref{fig5_kappa}; the event-plane alignment is examined separately in Appendix-\ref{app:angle}, where we show it reproduces the same pattern.

\begin{figure*}[t!]
    \centering
    \includegraphics[width=0.48\linewidth]{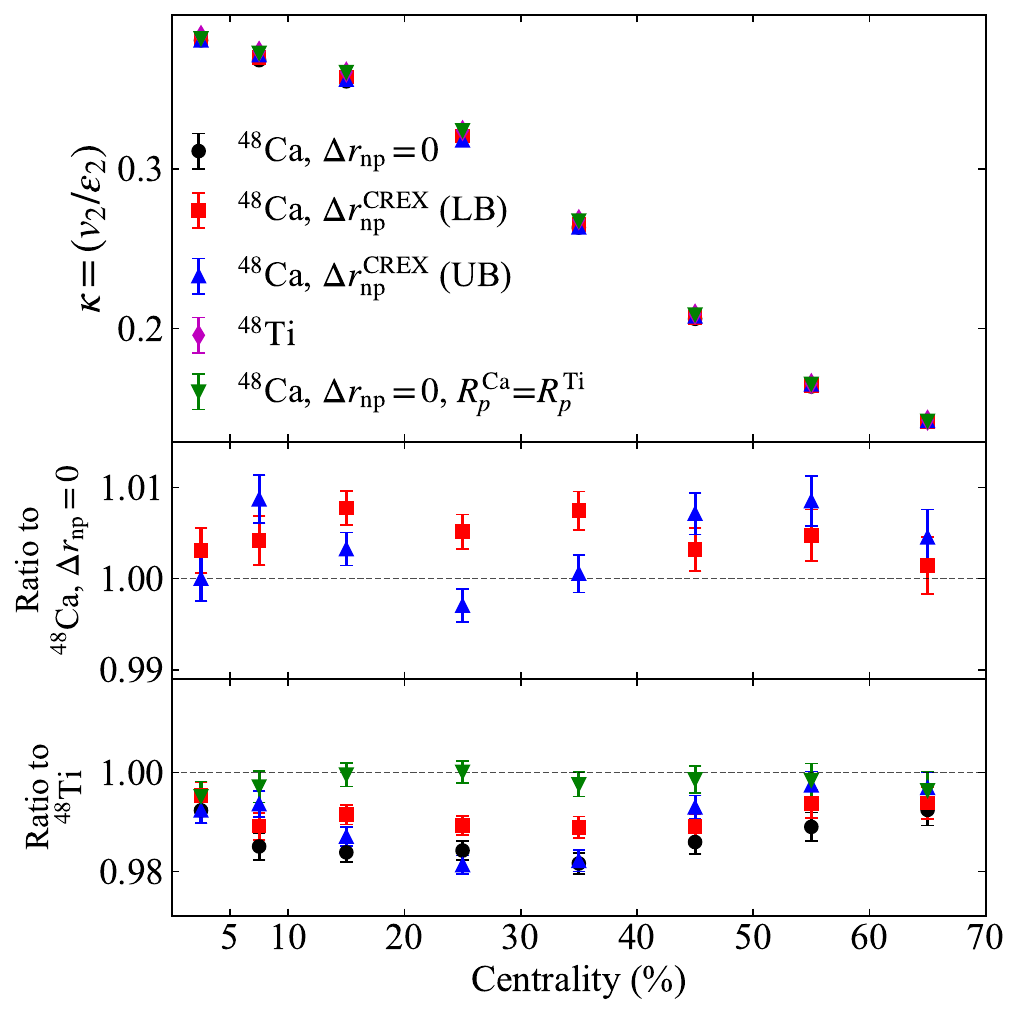}
    \includegraphics[width=0.48\linewidth]{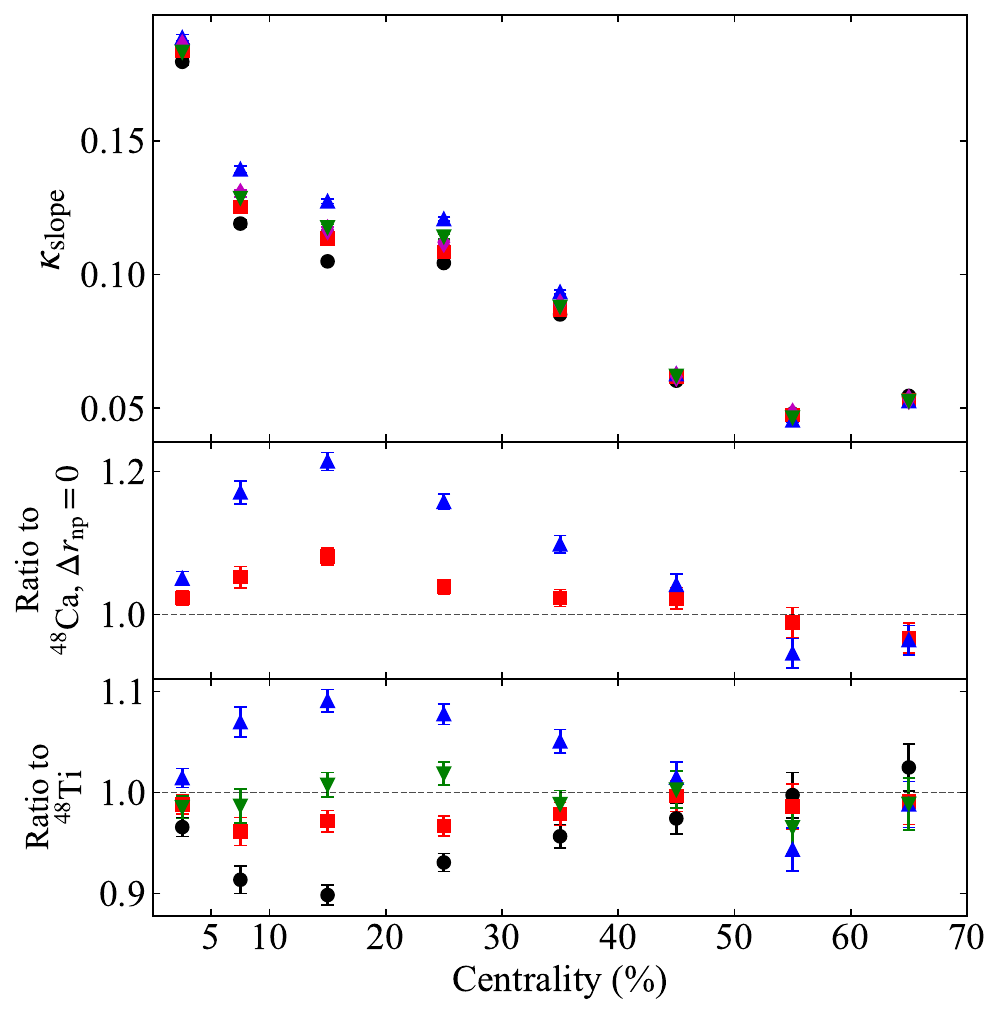}
    \includegraphics[width=0.48\linewidth]{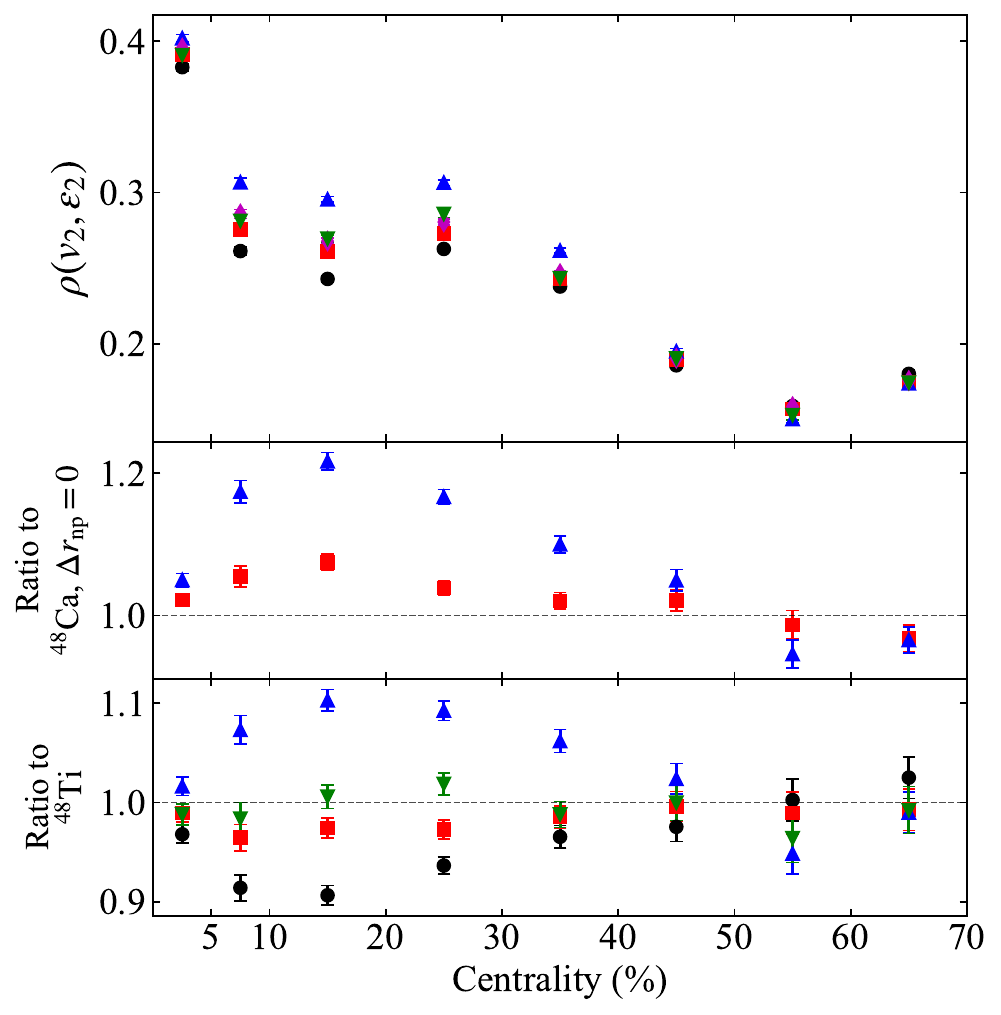}
    \begin{picture}(0,0)
        \put(-200,460){\bfseries(a)}
        \put(-20,460){\bfseries(b)}
        \put(-125,215){\bfseries(c)}
        \put(-10, 480){\fontsize{9}{10}\bfseries AMPT\(\boldsymbol{\sqrt{S_{NN}}=5.02}\) TeV}
        \put(-116, 232){\fontsize{9}{10}\bfseries \(\boldsymbol{0.2<p_T<3.0}\)}
    \end{picture}
    \caption{Centrality dependence of the mean linear response coefficient $\kappa=\langle v_2 \rangle / \langle \epsilon_2 \rangle$ (top left), event-by-event regression slope $\kappa_{\rm slope} = \mathrm{Cov}(v_2, \varepsilon_2)/\mathrm{Var}(\varepsilon_2)$ (top right) and the Pearson correlation $\rho(v_2, \varepsilon_2)$ (bottom), for the five nuclear configurations. Panel layout as in Fig. \ref{fig:fig1}. $\kappa$ is identical across all configurations to within $\sim 1\%$, confirming the medium's transport response is unaffected by the neutron skin; the same geometric input produces a $10–22\%$ $R_{\rm matter}$-ordered enhancement in $\kappa_{\rm slope}$ and $\rho(v_2,\varepsilon_2)$, showing that the skin's effect is confined to the event-by-event coupling between shape and flow.}
    \label{fig5_kappa}
\end{figure*}
Figure \ref{fig5_kappa}a presents the centrality dependence of $\kappa$ that falls from $\sim$0.38 in central collisions to $\sim$0.14 at 65\%, reflecting stronger viscous damping in smaller, shorter-lived peripheral systems. $\kappa_{\rm slope}$ (panel b) and $\rho(v_2,\varepsilon_2)$ (panel c) also decrease with centrality, as in large-nuclei collisions \cite{STAR:2015mki}. All five configurations lie on top of each other in $\kappa$ across the entire centrality range (Fig. \ref{fig5_kappa}a, top row). The skin changes it by $\lesssim$1\%, with no systematic centrality dependence (middle row). The ratio to $^{48}$Ti (bottom row) confirms every configuration, including the $R_p$-matched control, stays within $\sim 1-2\%$ of the $^{48}$Ti value. Such a small change is consistent with the parton-cascade and hadronization parameters being identical in all simulations. We conclude that the medium response is insensitive to the skin and depends only weakly on nuclear size.

Both $\kappa_{\rm slope}$ and $\rho(v_2,\varepsilon_2)$ are obtained from the event-by-event joint distribution of $v_2$ and $\varepsilon_2$. The linear regression $v_2 = \alpha + \kappa_{\rm slope}\varepsilon_2$ yields $\kappa_{\rm slope}$, and the normalized correlation strength is $\rho(v_2,\varepsilon_2) = \mathrm{Cov}(v_2,\varepsilon_2)/\sqrt{\mathrm{Var}(v_2)\mathrm{Var}(\varepsilon_2)}$. Both show a clear, $R_{\rm matter}$-ordered separation that appears already at 5-10\% centrality in top panels. 
Relative to the no-skin baseline, the neutron skin enhances $\kappa_{\rm slope}$ by $\sim$5-8\% for the CREX lower bound and $\sim$10-22\% for the CREX upper bound, an order of magnitude more than the enhancement of $\langle v_2\rangle$ at the same centrality. $\rho(v_2,\varepsilon_2)$ mathches the same enhancement like $\kappa_{\rm slope}$ peaking over the same 10-30\% centrality window, with the same $R_{\rm matter}$ ordering. 

The ratio to $^{48}$Ti is also larger. At 5–25\% centrality the no-skin baseline is 8-10\% below $^{48}$Ti in both observables. This is much larger than the deficits in any mean observable or in the variances. The $R_p$-matched control follows $^{48}$Ti to within $\sim$2\% across all centrality. The CREX LB lies $\sim$2–4\% below it and the UB $\sim$9-11\% above it in the 5–25\% range. Above 50\% the ordering is lost and all configurations, including $^{48}$Ti, lie close to each other.

Since $\kappa$ is essentially flat with $\Delta r_{np}$, the enhancement in $\kappa_{\rm slope}$ and $\rho(v_2,\varepsilon_2)$ is not a medium effect; it must reflect how the skin reshapes the joint distribution of $v_2$ and $\varepsilon_2$ event-by-event, not the mean mapping between them. The skin reshapes the geometric ensemble the medium processes, but leaves the medium's own response unchanged. The event-plane alignment $\langle\cos[2(\Psi_2^{\rm init}-\Psi_2^{\rm final})]\rangle$, an independent angular measure of the same event-by-event coupling, reproduces this pattern closely, with a $R_{\rm matter}$-ordered enhancement over a somewhat broader 5-60\% centrality range; we present the full analysis in Appendix-\ref{app:angle}. The two statistical interpretation of the joint distribution (a regression slope and a normalized correlation) together with an independent angular alignment measure all agree on centrality dependence, and $R_{\rm matter}$ ordering demonstrates that the neutron skin sharpens the initial participant geometry, which simultaneously tightens the magnitude and the angular coherence of the geometry-to-flow mapping. Because $\kappa$ itself is flat, none of this can be attributed to a modified medium response.

We note that we do not perform a systematic scan over parton cross section $\sigma_p$ in this work; since our claim is that the linear response coefficient $\kappa$ is insensitive to $\Delta r_{np}$, concerns a relative comparison across otherwise-identical simulations, we expect it to be robust to the specific
$\sigma_p$ value chosen, though we flag a dedicated sensitivity scan as a useful
check for future work.
With the medium response confirmed unmodified and the initial-state geometric origin of every flow enhancement now established, we turn next to the mixed shape–size correlators, which quantify how the skin coherently restructures both parameters of the overlap geometry together.

\subsection{Shape–size correlations and their final‑state signatures}
Sec. \ref{sec:medium} showed that the medium’s mean response is unchanged by the skin, while the event-by-event coupling between $v_2$ and $\varepsilon_2$ is enhanced. All observables considered so far treat eccentricity (shape) and inverse transverse size (size) as separate channels
The same nuclear density profile sources both, so they are are not independent \cite{Bozek:2016yoj} and a skin that modifies the overlap boundary should show up in their joint fluctuations \cite{Jia:2022qgl, Li:2019kkh, Bozek:2016yoj}. We quantify this with the covariance between the squared eccentricity and the relative fluctuation of the inverse transverse size,
\begin{equation}
    \mathrm{cov}(\varepsilon_2^2, \delta d_\perp/d_\perp) = \langle \varepsilon_2^2 \cdot \delta d_\perp/d_\perp\rangle - \langle \varepsilon_2^2 \rangle \langle \delta d_\perp/d_\perp\rangle
\end{equation}
where $\delta d_\perp/d_\perp= (d_\perp - \langle d_\perp\rangle)/\langle d_\perp\rangle$ is the relative event-by-event fluctuation of the inverse transverse size around its mean. The final-state counterpart is:
\begin{equation}
    \mathrm{cov}\big(v_2^2, \delta [p_T]/[p_T]\big) = \big\langle v_2^2 \cdot \delta [p_T]/[p_T]\big\rangle - \langle v_2^2 \rangle \langle \delta [p_T]/[p_T]\rangle
\end{equation}
with $\delta [p_T]/[p_T]= ([p_T]-\langle[p_T]\rangle)/\langle[p_T]\rangle$. The final-state covariance is computed with a
three-subevent method that mimics experimental non-flow suppression: subevents A
($-2.5<\eta_A<-0.75$), B ($-0.5<\eta_B<0.5$), and C ($0.75<\eta_C<2.5$). The quantity $v_2^2$ is estimated from the two‑particle correlation in the A–C pair using Eq. \ref{eq:v2cum}, while $\delta [p_T]/[p_T]$ is taken from the independent subevent B \cite{ATLAS:2022dov, Bally:2021qys}.

The corresponding normalized measure is the Pearson correlation coefficient \cite{Bally:2021qys, Giacalone:2020dln},

\begin{equation}
    \rho\big(v_2^2, \delta [p_T]/[p_T]\big) = \frac{\mathrm{cov}\big(v_2^2, \delta [p_T]/[p_T]\big)}{\sigma(v_2^2) \cdot \sigma(\delta [p_T]/[p_T])}
\end{equation}
If the response coefficients $v_2/\varepsilon_2$ and $[p_T]/d_\perp$ do not depend on the skin (as established in \ref{sec:medium}), these final-state covariances map directly onto their initial-state counterparts, making them experimentally accessible probes of the same combined geometric restructuring of the overlap zone.\\

\begin{figure*}[t!]
    \centering
    \includegraphics[width=0.48\linewidth]{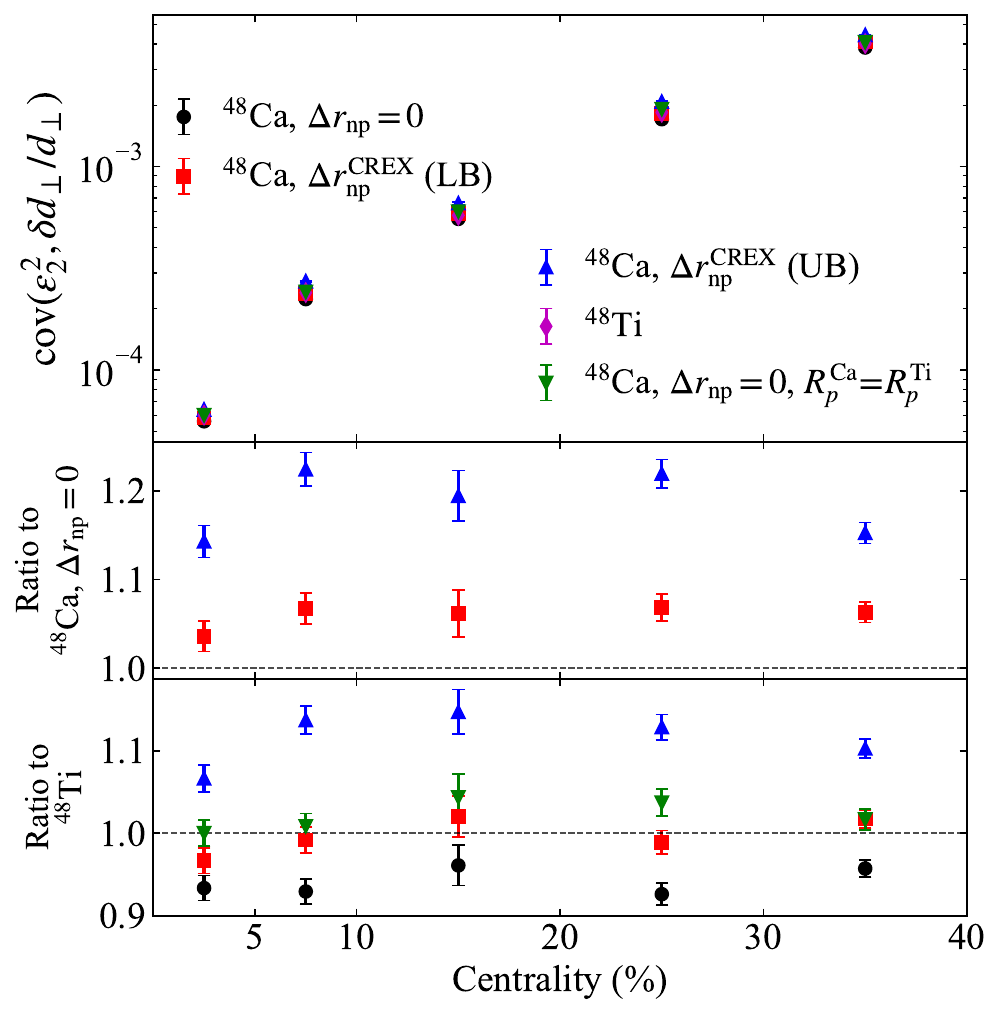}
    \includegraphics[width=0.48\linewidth]{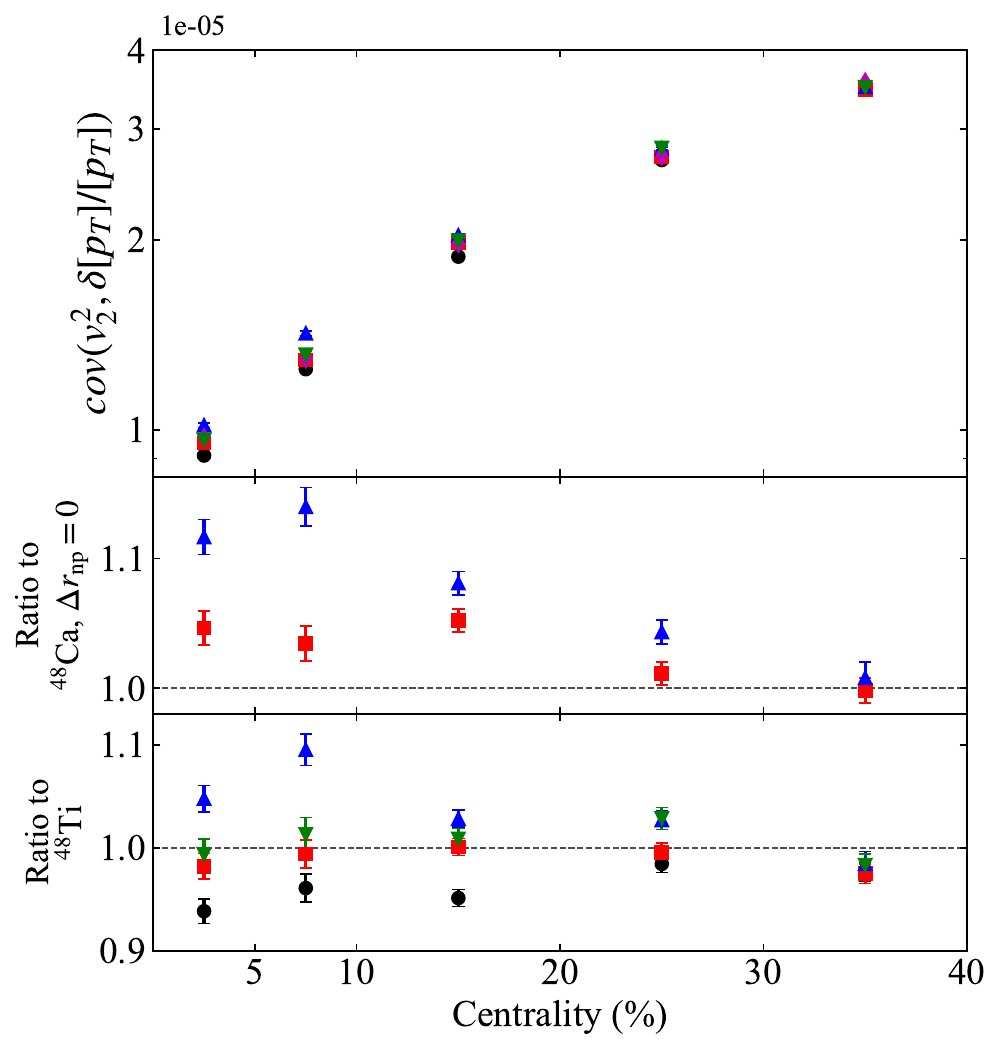}
    \begin{picture}(0,0)
        \put(-450,230){\bfseries (a)}
        \put(-205,230){\bfseries (b)}
        \put(-180, 230){\fontsize{8}{10}\bfseries AMPT\(\boldsymbol{\sqrt{S_{NN}}=5.02}\) TeV}
        \put(-110, 150){\fontsize{8}{10}\bfseries \(\boldsymbol{0.2<p_T<3.0}\)}
    \end{picture}
    \caption{Centrality dependence of the initial-state shape–size covariance $\mathrm{cov}(\varepsilon_2^2, \delta d_\perp/d_\perp)$ (left) and its final-state counterpart $\mathrm{cov}(v_2^2, \delta [p_T]/[p_T])$ (right), for the five nuclear configurations. Panel layout as in Fig. \ref{fig:fig1}. The skin's enhancement of the covariance (15–25\% initial state, 9–12\% final state) exceeds the enhancement seen in either variance alone (Figs. \ref{fig2:var_dp_mpt}, \ref{fig4_var_eps2_v2}), showing that the skin couples the shape and size fluctuations rather than shifting them independently.}
    \label{fig6_cov}
\end{figure*}

Figure \ref{fig6_cov} shows the initial-state covariance $\mathrm{cov}(\varepsilon_2^2, \delta d_\perp/d_\perp)$ (left) and final state covariance  $\mathrm{cov}(v_2^2, \delta [p_T]/[p_T])$ (right) as a function of centrality. Both rise monotonically with centrality, similar to what is seen for deformed ($^{238}$U) and spherical ($^{197}$Au) nuclei \cite{Jia:2021qyu}.The initial-state covariance rises from $\sim6\times10^{-5}$ to $\sim4\times10^{-3}$, and the final-state one from $\sim10^{-5}$ to $\sim3.5\times10^{-5}$.


In the initial state, the skin enhances $\mathrm{cov}(\varepsilon_2^2,\delta d_\perp/d_\perp)$ by $\sim$3–7\% for the lower bound and $\sim$14–22\% for the upper bound, with little centrality dependence (Fig. \ref{fig6_cov}a). This is larger than the change of any individual second moment in Figs. \ref{fig2:var_dp_mpt} and \ref{fig4_var_eps2_v2}. This is the clearest evidence that the skin does not shift the shape and size distributions independently. A sharper nuclear edge makes eccentric-and-compact events occur together more often, tightening the correlation between the two geometric features. In the isobaric comparison, the no-skin configuration lies $\sim$6–8\% below $^{48}$Ti and the LB within $\sim$3\% of it. The UB lies 8–15\% above it. The $R_p$-matched control is very close to the $^{48}$Ti baseline, and the same $R_{\rm matter}$ control established earlier, now shown to hold for a joint correlator as well as for individual moments.

The final-state covariance shows the same ordering with a smaller amplitude (Fig. \ref{fig6_cov}b). The UB enhancement is $\sim$10–12\% in central collisions and the LB $\sim$5\%, and both vanish by 35\% centrality.
This attenuation is qualitatively consistent with medium expansion smoothing fluctuation-driven correlations, but we have not propagated the measured response coefficients to a quantitative prediction of the attenuated covariance, and we note this as a natural target for future work. The isobaric comparison shows that the no-skin baseline lies 5–8\% below $^{48}$Ti and the LB $\sim 2$\% below it in central collisions, while the matched control follows within $\sim$1\% of the $^{48}$Ti and the upper bound lies 5-10\% above it.

\begin{figure}
    \centering
    \includegraphics[width=\linewidth]{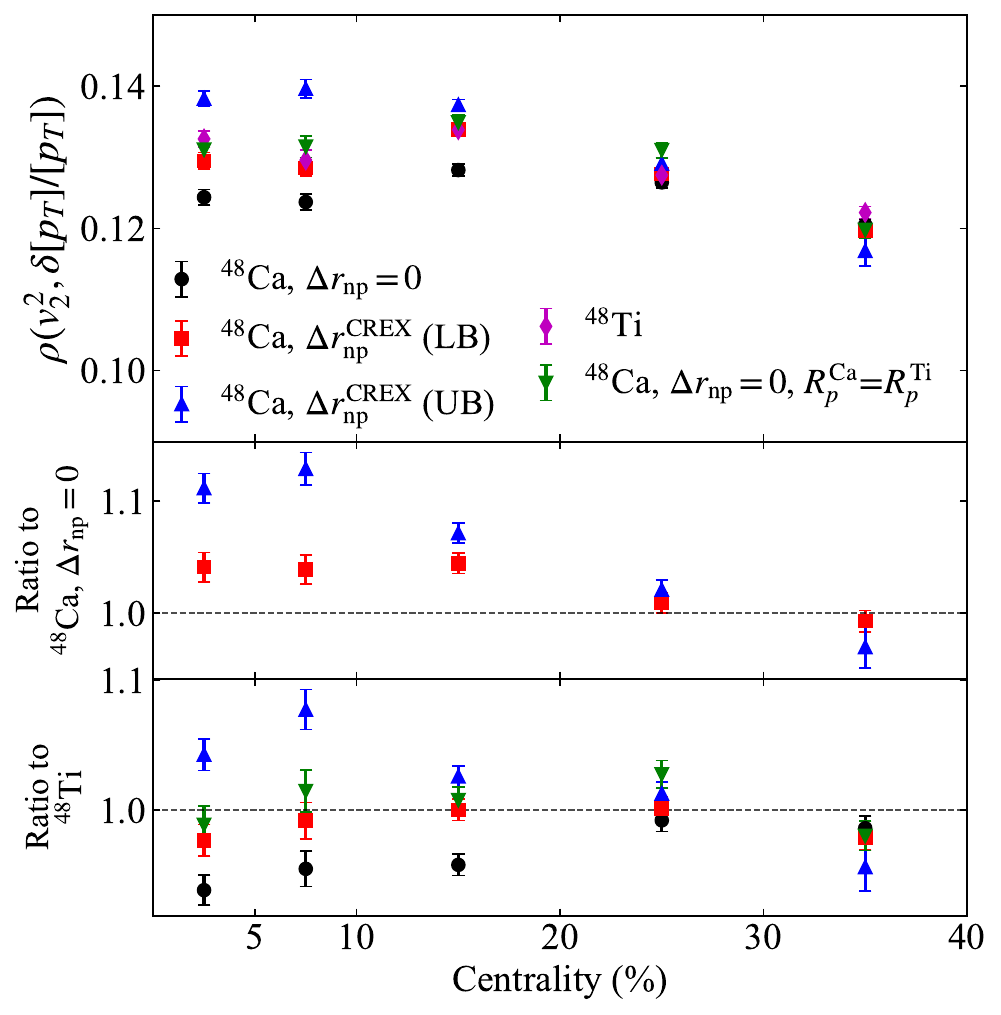}
    \begin{picture}(0,0)
        \put(-10, 243){\fontsize{8}{10}\bfseries AMPT\(\boldsymbol{\sqrt{S_{NN}}=5.02}\) TeV}
    \end{picture}
    \caption{Centrality dependence of the Pearson correlation coefficient $\rho(v_2^2, \delta [p_T]/[p_T])$ between event-by-event $v_2^2$ and mean-$[p_T]$ fluctuations, for the five nuclear configurations. Panel layout as in Fig. \ref{fig:fig1}. Unlike the monotonic covariance of Fig. \ref{fig6_cov}, $\rho$ peaks at 5–10\% centrality and decreases on both sides; normalizing by the individual variances also removes the broad enhancement in centrality and isolates a more centrally localized 8–11\% skin signature.}
    \label{fig7_rho2}
\end{figure}

Figure \ref{fig7_rho2} shows the Pearson correlation $\rho(v_2^2, \delta [p_T]/[p_T])$, which removes the individual variances from Fig. \ref{fig6_cov}. In contrast to the covariance, $\rho$ is non-monotonic. It peaks at 5–15\% and decreases toward 35\%. Earlier studies found only a weak centrality dependence for spherical nuclei \cite{Schenke:2020uqq}. This non-monotonic shape reflects a covariance that grows with centrality, while the product of standard deviations in the denominator grows at a comparable or faster rate toward peripheral centralities.
The skin enhances $\rho(v_2^2, \delta [p_T]/[p_T])$ by up to $\sim 12\%$ for CREX upper bound and by $\sim 4-5\%$ for CREX lower bound relative to the no-skin baseline in the 0-20\% centrality range, converging to unity by 30\% centrality. The isobaric comparison follows Fig. \ref{fig6_cov}b. The no-skin baseline is 5-6\% below $^{48}$Ti at 0–15\%, the LB is within $\sim$2\% of it, the UB is 4–7\% above it, and the matched control follows the baseline to within $\sim$1-2\%.\\

The simultaneous enhancement of $\rho(v_2^2, \delta [p_T]/[p_T])$ and of the second-moment geometry-flow coupling measures ($\kappa_{\rm slope}$, $\rho(v_2, \varepsilon_2)$, event-plane alignment) established in the previous section (Sec. \ref{sec:medium}) completes a coherent picture. The neutron skin sharpens the initial participant geometry, which simultaneously tightens the correlation between shape and size, and between eccentricity and final flow. The covariance and its normalized form together constitute the most sensitive and physically transparent observables in this analysis, encoding the full correlated geometric response of the overlap region to the neutron skin in a single, experimentally accessible quantity, the natural target for a future $^{48}$Ca+$^{48}$Ca measurement at the LHC.  We note that resolving second-moment and covariance observables at the percent-level precision indicated above requires substantially larger event samples than are needed for mean observables, since statistical uncertainties on variances and covariances scale less favorably with event count. A quantitative projection of the integrated luminosity or minimum-bias event count required for a $5\sigma$ discrimination between the CREX bounds is beyond the scope of this study, but should be a priority for follow-up work assessing the experimental feasibility of this program at the LHC.

\section{Summary and Conclusion} \label{secIV}
We have presented a systematic study of neutron skin effects in $^{48}$Ca+$^{48}$Ca at $\sqrt{s_{NN}} = 5.02$ TeV using AMPT simulations with three nuclear density profiles: a zero-skin baseline and skins at the lower and upper CREX bounds of $\Delta r_{np}$. To separate the skin from the nuclear size, we introduced $^{48}$Ti+$^{48}$Ti and a radius-matched, skin-free $^{48}$Ca configuration as reference systems. This separates the two effects to within the residual offsets of the matched control, which reach a few percent (up to 2\%) in some observables and are quantified below.

Our results indicate that the neutron skin acts through the initial-state geometry. The mean linear response $\kappa=\langle v_2\rangle/\langle\varepsilon_2\rangle$ changes by $\lesssim 1\%$ with the skin, with no systematic trend, while the event-by-event regression slope $\kappa_{\rm slope}$ is enhanced by up to 22\% for the CREX upper bound below 45\% centrality. The correlation $\rho(v_2,\varepsilon_2)$ and the event-plane alignment (up to $\sim$12\%; shown in Appendix~\ref{app:angle}) show the same enhancement. Since the mean response is flat, we find no evidence that these enhancements reflect a change of the medium response. The skin reshapes the geometry the medium processes. A change of the overall nuclear size, at fixed skin, also changes $\kappa$ by only $\sim$1\%. We did not scan the parton cross section, so the origin of the tighter $v_2$--$\varepsilon_2$ correlation remains open.

The position of $^{48}$Ti relative to the skin configurations depends on the radius assigned to it, and this is the essential consideration for any Ca/Ti comparison. With $R_p=R_n=3.515$ fm (main text), the experimental rms point-proton radius used directly as the Woods-Saxon radius, the matter rms radius of $^{48}$Ti lies between those of the CREX lower and upper bounds. With the larger $R_p=R_n=3.56$ fm (Appendix \ref{app:ti}), which reproduces the experimental rms point-proton radius in the sampled distribution, it lies close to that of the CREX upper bound. The same $^{48}$Ca skin configuration therefore appears thicker or thinner than $^{48}$Ti depending on which $^{48}$Ti is chosen, and the Ca/Ti ratio changes accordingly. The radius-matched $^{48}$Ca follows $^{48}$Ti to within $\sim$1\% in the mean observables, but only to $\sim$1-2\% in several fluctuation and correlation observables, including the shape-size covariance. A comparison with $^{48}$Ti alone is therefore not sufficient evidence of skin sensitivity; the size of $^{48}$Ti has to be controlled.

The sensitivity to $\Delta r_{np}$ grows from means to fluctuations to joint correlations. The mean observables show only few-percent sensitivity. The variances reach $\sim 5-8\%$ for the upper bound, although part of the change in those observables comes from the rescaling. The coupling observables ($\kappa_{\rm slope}$, $\rho(v_2,\varepsilon_2)$, event-plane alignment) and the shape-size covariance reach 10-22\% in the initial state, several times the mean-level effect. These require the initial-state eccentricity $\varepsilon_2$ or the initial event plane $\Psi_2^{\rm init}$ and cannot be measured directly. Their measurable final-state counterparts, $\mathrm{cov}(v_2^2,\delta[p_T]/[p_T])$ and the Pearson coefficient $\rho(v_2^2,\delta[p_T]/[p_T])$, show enhancements of $\sim$10-12\% for the upper bound. These are the most sensitive measurable observables we identified, but the enhancements decrease toward 35\% centrality. This may be due to centrality-bin migration and area inflation at fixed $N_{\rm part}$, but we have not tested this. For a future run we suggest $\mathrm{cov}(v_2^2,\delta[p_T]/[p_T])$ and $\rho(v_2^2,\delta[p_T]/[p_T])$, together with $v_2\{2\}$, interpreted against both $^{48}$Ti and the radius-matched control.

A few limitations of this study should be noted. The results come from one AMPT parameter set with a fixed diffuseness parameter, and they include no short-range correlations or clustering and no sub-nucleonic fluctuations. Following Ref. \cite{Vitsos:2025jzt}, we use experimental rms point-proton radii directly as half-density radii, so the sampled rms radii differ from the experimental ones by $+0.023$ fm for $^{48}$Ca and, in the main-text setting, $-0.028$ fm for $^{48}$Ti. These offsets are of the same order as the 0.01 fm level at which the reference radius must be controlled, so a setup that matches the sampled rms of both nuclei to experiment is a necessary refinement. We also neglect the small neutron skin of $^{48}$Ti. Finally, resolving second-moment and covariance observables at the percent level requires larger event samples than for means, and we have not estimated the number of events needed to separate the CREX bounds; this is essential for assessing the feasibility of a measurement.

In summary, the clearest heavy-ion signature of the neutron skin in our simulations lies not in mean flow but in second-moment and joint-correlation observables. This offers a route to constrain the neutron skin of $^{48}$Ca at the LHC that is independent of, and complementary to, CREX, provided the size of the reference nucleus is controlled. The framework can also be extended to $^{208}$Pb and $^{124}$Sn \cite{Hyun:2021osj, Zenihiro:2010zz, Tarbert:2013jze, Terashima:2008zza} and to other near-isobaric pairs such as $^{96}$Zr/$^{96}$Ru, where a size-matched control can separate skin from size and deformation. With realistic initial-state fluctuations, higher-order harmonics and a Bayesian analysis of $\Delta r_{np}$ using many observables, heavy-ion collisions could provide a determination of the $^{48}$Ca skin that complements direct scattering experiments. 


\bibliography{ref.bib}

\appendix
\section{Event-plane alignment}
\label{app:angle}

\begin{figure*}
    \centering
    \includegraphics[width=0.48\linewidth]{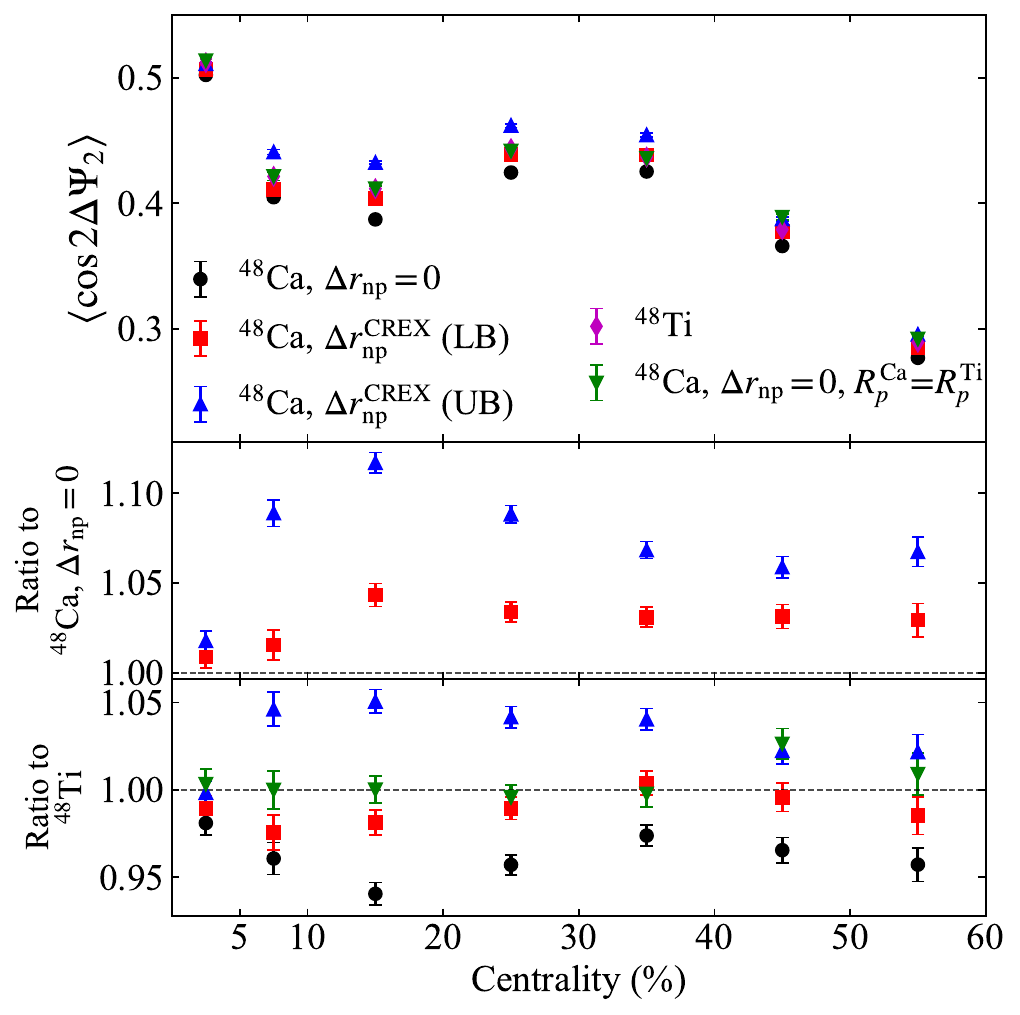}
    \begin{picture}(0,0)
        \put(-130, 230){\bfseries AMPT\(\boldsymbol{\sqrt{S_{NN}}=5.02}\) TeV}
    \end{picture}
    \caption{Centrality dependence of the event-plane alignment $\langle \cos[2(\Psi_2^{\rm init}- \Psi_2^{\rm final})]\rangle$, for the five nuclear configurations. Panel layout as in Fig. \ref{fig:fig1}. The alignment reproduces the $R_{\rm matter}$-ordered enhancement of Fig. \ref{fig5_kappa} over a broader centrality range, confirming that the skin's effect on the event-by-event geometry-flow coupling is not an artifact of a single statistical measure.}
    \label{figA_psi}
\end{figure*}
As a further test of the event-by-event geometry-to-flow coupling discussed in Sec. \ref{sec:medium}, we examine the event-plane alignment $\langle\cos[2( \Psi_2^{\rm init}-\Psi_2^{\rm final})]\rangle$.
This is an angular measure of the coupling, whereas $\kappa_{\rm slope}$ and $\rho(v_2,\varepsilon_2)$ are built from the magnitudes of $v_2$ and $\varepsilon_2$.
Unlike $\kappa_{\rm slope}$ and $\rho(v_2,\varepsilon_2)$, which are derived from a single per-event regression or covariance, the alignment correlator integrates over the full flow-vector orientation, including contributions from spectator-driven plane definition and longer-range geometric correlations \cite{Adams:2021yob, Xiao:2012uw}. It is also sensitive to the decorrelation of the participant and flow planes.

Figure \ref{figA_psi} shows the centrality dependence of the event-plane alignment for the five nuclear configurations. The skin-induced enhancement is comparatively smaller but qualitatively consistent with the $\kappa_{\rm slope}$ and $\rho(v_2,\varepsilon_2)$ pattern of Fig. \ref{fig5_kappa}. It falls from $\sim$0.5 in the most central bin to $\sim$0.4 at 7.5–15\%, has a local maximum at 25–35\%, and decreases to $\sim$0.28 at 55\%. Relative to the no-skin baseline, the skin enhances the event-plane alignment by $2$-$5\%$ for the CREX lower bound and $8$-$12\%$ for the CREX upper bound. The UB enhancement peaks at 15\% and stays at $\sim$6–7\% from 35\% to 55\%.

In the isobaric comparison, the no-skin configuration falls $4$-$6\%$ below $^{48}$Ti, with the largest deficit at 15\%. The LB is within $\sim$2.5\% of $^{48}$Ti and the UB is $\sim$5\% above it. The $R_p$-matched control follows $^{48}$Ti to within $\sim$1\% up to 35\% with the same $R_{\rm matter}$-ordered pattern established for every other coupling observable in this study. This independent angular measure therefore confirms that the enhancement seen in $\kappa_{\rm slope}$ and $\rho(v_2,\varepsilon_2)$ is not an artifact of either statistical measure chosen.


\section{Dependence on the $^{48}$Ti radius}
\label{app:ti}

In the main text, $^{48}$Ti and the radius-matched $^{48}$Ca (Case 5) use $R_p=R_n=3.515$ fm, i.e. the experimental rms point-proton radius of $^{48}$Ti \cite{Wu:2025sgt} used directly as the half-density radius, in the same way as for $^{48}$Ca. Because $R$ is not an rms radius, the sampled rms of this configuration is 3.487 fm, below the experimental value. To test how the Ca/Ti comparison depends on the size of the reference, we repeat Cases 4 and 5 with $R_p=R_n=3.56$ fm. For $a=0.586$ fm this gives a sampled rms of 3.514 fm, which reproduces the experimental rms point-proton radius of $^{48}$Ti. Cases 1-3 are unchanged. Table~\ref{tab:tiapp} compares the two settings.

\begin{table*}[htbp]
\centering
\caption{The two $^{48}$Ti settings. $R_p=R_n$ is the Woods--Saxon half-density radius ($a=0.586$ fm) and $\langle r^2\rangle^{1/2}$ is the rms radius of the sampled distribution. The last row is the difference between the experimental point-proton radii of $^{48}$Ti and $^{48}$Ca (3.387 fm).}
\begin{tabular}{l c c c c c}
\toprule
Setting & $R_p$ (fm) & $R_n$ (fm) & $\langle r^2\rangle^{1/2}$ (fm) & $\langle r^2\rangle^{1/2}-\langle r^2\rangle^{1/2}_{\rm Case\,1}$ (fm)\\
\midrule
Main text (case 4 \& 5) & 3.515 & 3.515 & 3.487 & 0.077 \\
Appendix (case 4 \& 5) & 3.56  & 3.56  & 3.514 & 0.104 \\
Experiment (point protons) & --- & --- & 3.515 & 0.128 \\
\bottomrule
\end{tabular}
\label{tab:tiapp}
\end{table*}

In the main-text setting the matter rms of $^{48}$Ti lies between those of the CREX lower and upper bounds (3.450 and 3.517 fm). In the setting used here it coincides with the upper bound to within 0.003 fm. The two settings therefore differ in where $^{48}$Ti sits relative to the CREX range, and this changes how the Ca/Ti ratios are read. The experimental rms difference between $^{48}$Ti and $^{48}$Ca is 0.128 fm, and the setting used here reproduces 0.104 fm of it.

\begin{figure*}[htbp]
\centering
\includegraphics[width=0.49\textwidth]{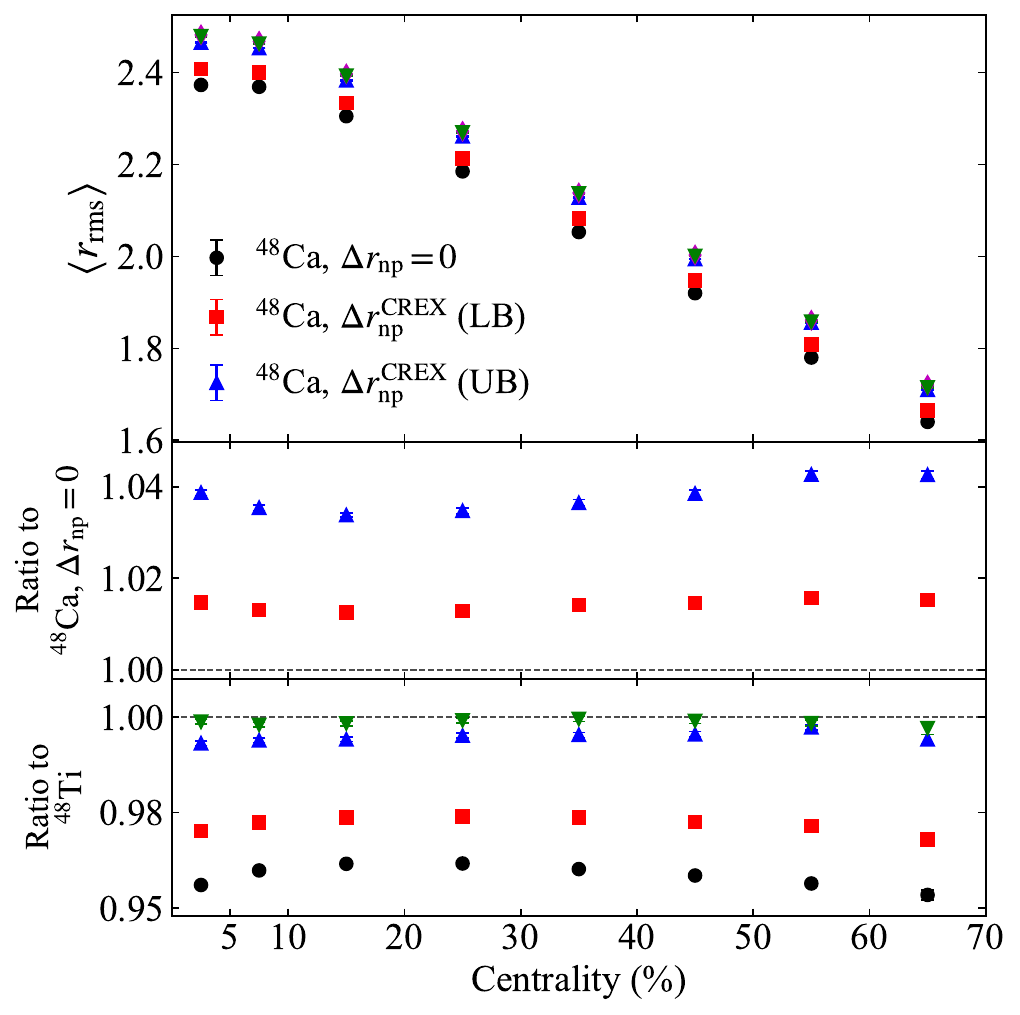}
\includegraphics[width=0.49\textwidth]{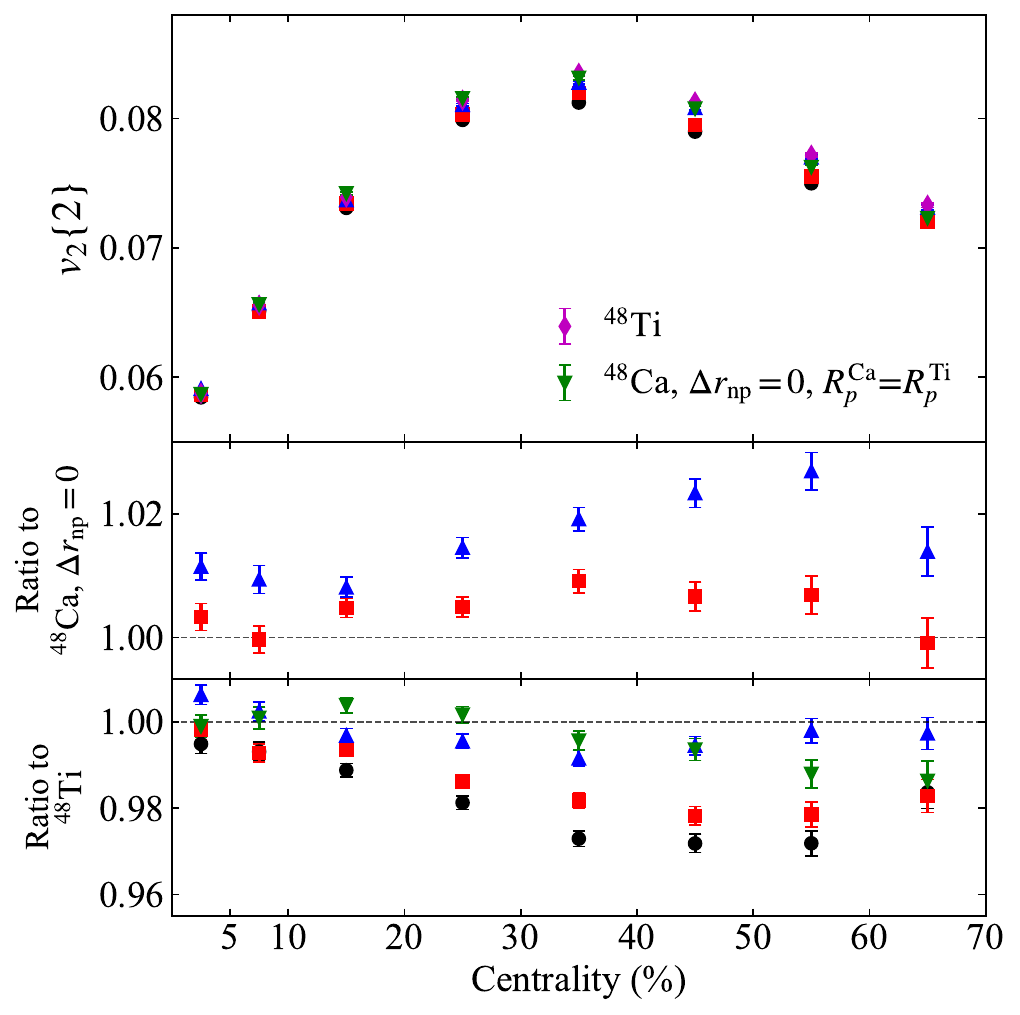}
\includegraphics[width=0.49\textwidth]{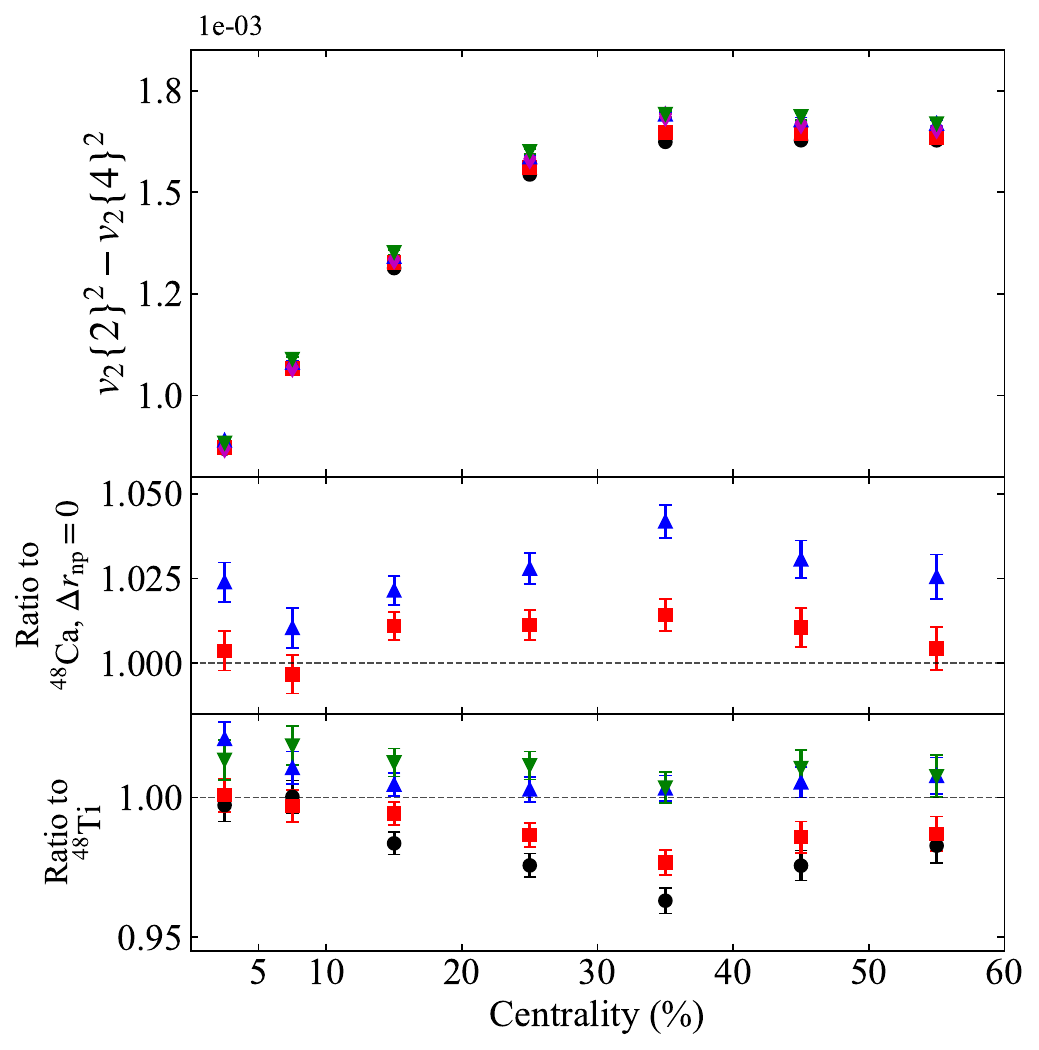}
\begin{picture}(0,0)
    \put(-320,460){\bfseries (a)}
    \put(-75,460){\bfseries (b)}
    \put(-195,210){\bfseries (c)}
    \put(-260, 485){\fontsize{9}{10}\bfseries AMPT\(\boldsymbol{\sqrt{S_{NN}}=5.02}\) TeV}
    \put(22, 485){\fontsize{9}{10}\bfseries \(\boldsymbol{0.2<p_T<3.0}\)}
\end{picture}
\caption{Same as Figs. \ref{fig:fig1}a, \ref{fig3_eps2_v2}b and \ref{fig4_var_eps2_v2}b, but with $R_p=R_n=3.56$ fm for $^{48}$Ti and for the radius-matched $^{48}$Ca: $\langle r_{\rm rms}\rangle$ (panel a), $v_2\{2\}$ (panel b) and $\mathrm{Var}(v_2)$ (panel c). Panel layout as in Fig. \ref{fig:fig1}.}
\label{fig:app1}
\end{figure*}

\begin{figure*}[htbp]
\centering
\includegraphics[width=0.49\textwidth]{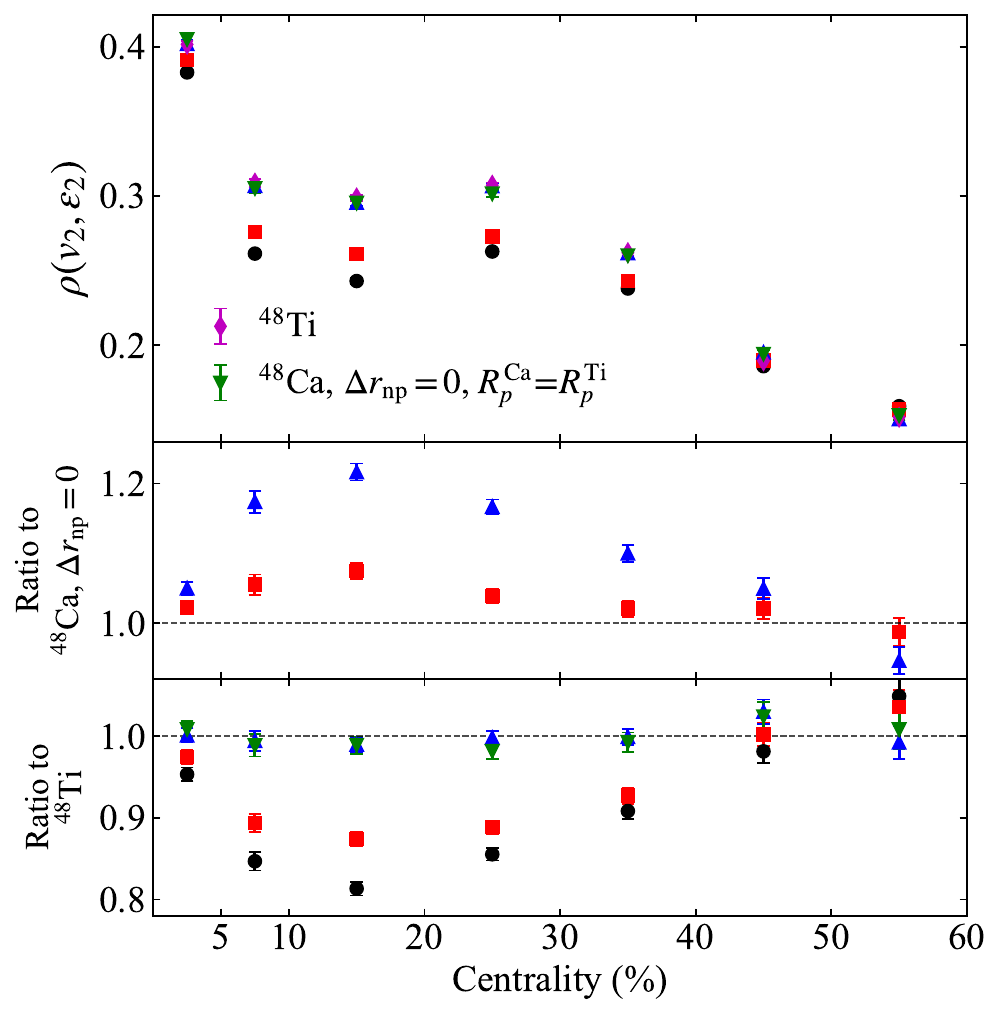}
\includegraphics[width=0.49\textwidth]{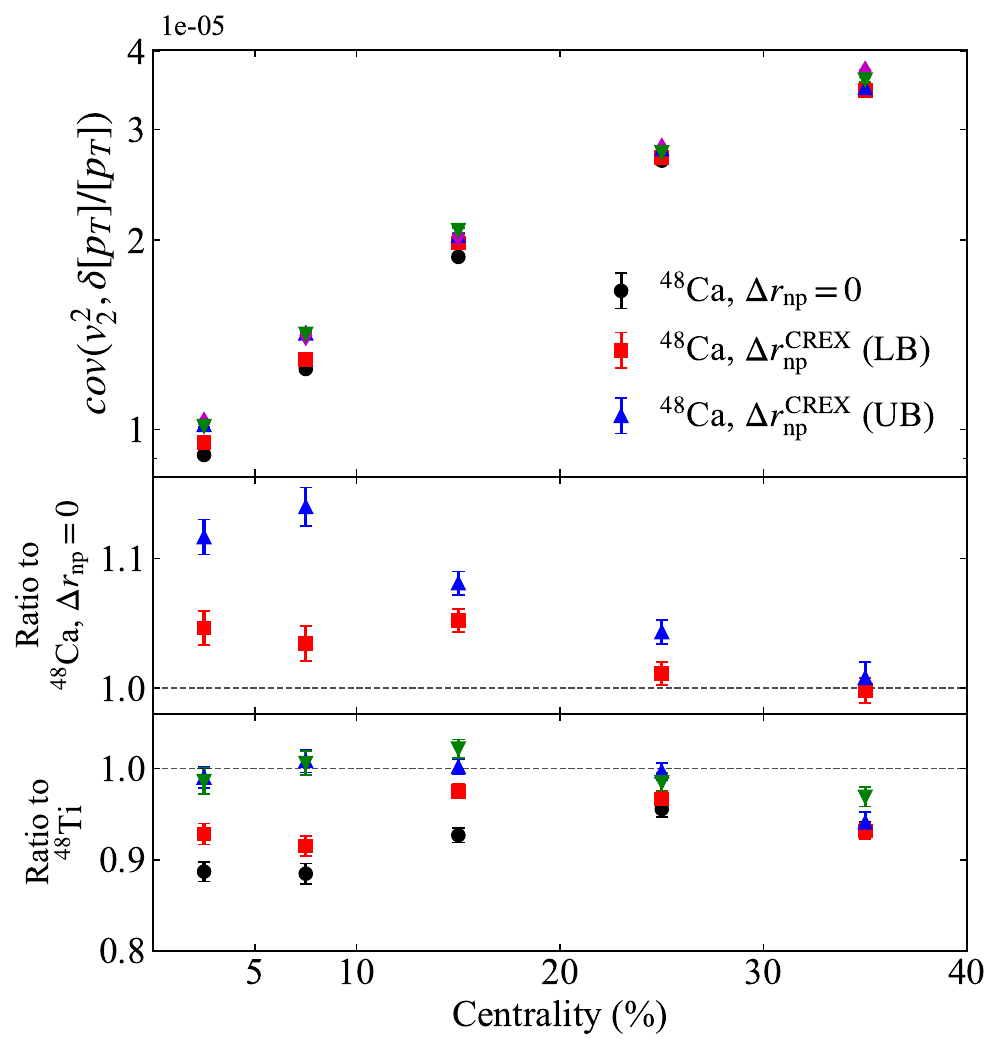}
\includegraphics[width=0.49\textwidth]{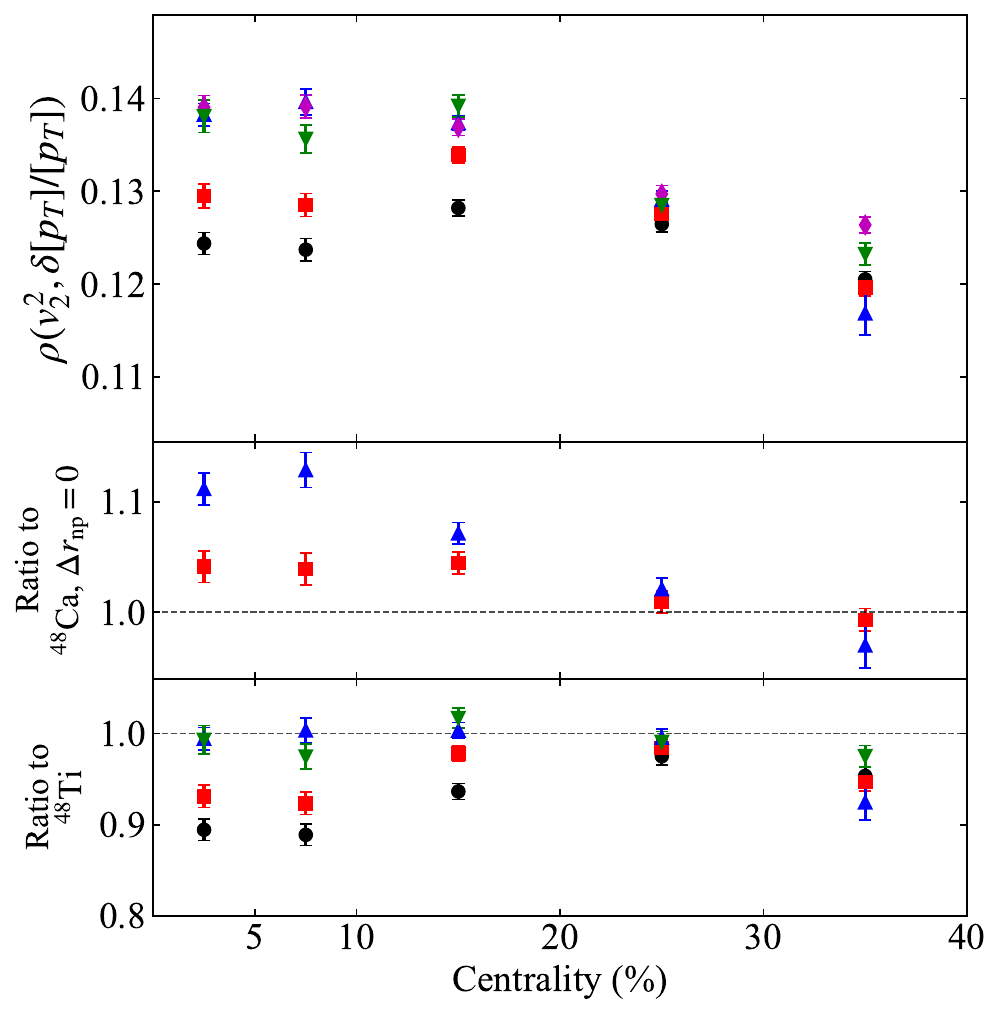}
\begin{picture}(0,0)
    \put(-300,490){\bfseries (a)}
    \put(-75,490){\bfseries (b)}
    \put(-200,240){\bfseries (c)}
    \put(-260, 490){\fontsize{9}{10}\bfseries AMPT\(\boldsymbol{\sqrt{S_{NN}}=5.02}\) TeV}
    \put(-42, 490){\fontsize{9}{10}\bfseries \(\boldsymbol{0.2<p_T<3.0}\)}
\end{picture}
\caption{Same as Figs. \ref{fig5_kappa}c, \ref{fig6_cov}b and \ref{fig7_rho2}, but with $R_p=R_n=3.56$ fm for $^{48}$Ti and for the radius-matched $^{48}$Ca: $\rho(v_2,\varepsilon_2)$ (panel a), $\mathrm{cov}(v_2^2,\delta[p_T]/[p_T])$ (panel b) and $\rho(v_2^2,\delta[p_T]/[p_T])$ (panel c). Panel layout as in Fig. \ref{fig:fig1}.}
\label{fig:app2}
\end{figure*}

Figure \ref{fig:app1} show $\langle r_{\rm rms}\rangle$, $v_2\{2\}$, $\mathrm{Var}(v_2)$, and Figure \ref{fig:app2} show $\rho(v_2,\varepsilon_2)$, $\mathrm{cov}(v_2^2,\delta[p_T]/[p_T])$ and $\rho(v_2^2,\delta[p_T]/[p_T])$ with the larger $^{48}$Ti radius. The middle rows are identical to the corresponding figures of Sec. \ref{secIII}, since they involve only Cases 1-3, and we do not discuss them again. The bottom rows change, and they are what we compare here.

For $\langle r_{\rm rms}\rangle$ the no-skin baseline lies $\approx4-5\%$ below $^{48}$Ti, the lower bound $\approx2-2.5\%$ below, and the upper bound agrees with $^{48}$Ti to within $\approx0.5\%$. The radius-matched control reproduces $^{48}$Ti within the statistical uncertainty. Since $^{48}$Ti now sits at the upper-bound size, the ratio crosses unity at the upper bound, while in the main text it crosses between the two bounds.

For $v_2\{2\}$ the no-skin baseline lies 1-3\% below $^{48}$Ti, with the deficit growing toward peripheral collisions, and the upper bound is within $\approx1\%$. For $\mathrm{Var}(v_2)$ the no-skin baseline is $\approx5\%$ below $^{48}$Ti at 30-40\% centrality and the upper bound is again the closest. The matched control follows $^{48}$Ti in both observables.

The largest effects are in $\rho(v_2,\varepsilon_2)$ and in the final-state shape--size observables in Fig. \ref{fig:app2}. The no-skin baseline lies 15-20\% below $^{48}$Ti in $\rho(v_2,\varepsilon_2)$ at 5-30\% centrality, the lower bound $\approx12\%$ below, and the upper bound agrees with $^{48}$Ti within $\approx1\%$. For $\mathrm{cov}(v_2^2,\delta[p_T]/[p_T])$ and $\rho(v_2^2,\delta[p_T]/[p_T])$ the no-skin baseline is 10-13\% below $^{48}$Ti in the central collisions, the upper bound is within 2-4\%, and the matched control is within 1-2\%.

Going from $R_p=R_n=3.515$ fm to 3.56 fm increases the sampled rms radius of $^{48}$Ti and of the radius-matched $^{48}$Ca from 3.487 to 3.514 fm. AS cases 1-3 are unchanged, every comparison that involves only the $^{48}$Ca skin configurations is the same as in the main text. The change appears only in the ratios to $^{48}$Ti, and mostly in how the three skin configurations are ordered relative to it.

In the main text the matter rms radius of $^{48}$Ti lies between those of the lower and upper bound, and the ratios to $^{48}$Ti cross unity somewhere between the two. With the larger radius, $^{48}$Ti is about as large as the upper-bound configuration (3.514 against 3.517 fm). The upper-bound curve then moves close to unity, and the lower bound and the no-skin baseline move further below $^{48}$Ti. For $\rho(v_2,\varepsilon_2)$ and for the final-state covariance and its Pearson coefficient the shifts are larger, because these observables depend more strongly on the system size.

Two features do not change. The ordering of the configurations by size, i.e., no-skin, lower bound, upper bound, is the same in both settings. The radius-matched $^{48}$Ca follows $^{48}$Ti to the same precision as in the main text, as it must, since the two configurations have identical density parameters and differ only in $Z/N$.

Taken together, the two settings show how much of the Ca/Ti comparison is set by the size assigned to $^{48}$Ti. A change of 0.027 fm in the sampled rms radius of the reference moves the ratios to one side of the $^{48}$Ti baseline. The skin effect measured against the zero-skin baseline does not depend on this choice. A Ca/Ti comparison used to constrain $\Delta r_{np}$ therefore needs the size of $^{48}$Ti to be fixed from experiment, and our two settings, which differ from the experimental rms difference between $^{48}$Ti and $^{48}$Ca (0.128 fm) by 0.051 and 0.024 fm, bracket the uncertainty that this adds.
\end{document}